\documentstyle[prl,aps,mypsfig2,twocolumn,floats]{revtex}
\begin{document}

\preprint{twobit.tex}

\title{Experimental Realization of A Two Bit Phase Damping Quantum Code}

\author{Debbie Leung$^{1,2}$, 
	Lieven Vandersypen$^{3,2}$,
	Xinlan Zhou$^{3,2}$,
	Mark Sherwood$^{2}$,
\\
	Constantino Yannoni$^{2}$,
	Mark Kubinec$^{4}$,
	Isaac Chuang$^{2,3}$
}

\address{\vspace*{1.2ex}
	{$^1$ Edward L. Ginzton Laboratory, Stanford University  
		 \\ Stanford, CA 94305-4085}\\[1.2ex]
 	{$^2$ IBM Almaden Research Center \\ San Jose, CA
 	94120}\\[1.2ex]
	{$^3$ Solid State and Photonics Laboratory, Stanford University 
		 \\ Stanford, CA 94305-4075}\\[1.2ex]
        {$^4$    College of Chemistry, D7 Latimer Hall, \\
                University of California Berkeley, Berkeley, CA 94720-1460}
}
	
\date{\today}
\maketitle

%%%%%%%%%%%%%%%%%%%%%%%%%%%%%%%%%%%%%%%%%%%%%%%%%%%%%%%%%%%%%%%%%%%%%%%%%%%%%
% Abstract

% \begin{onecolumn}

\begin{abstract}
Using nuclear magnetic resonance techniques, we experimentally investigated
the effects of applying a two bit phase error detection code to preserve
quantum information in nuclear spin systems.  Input states were stored with
and without coding, and the resulting output states were compared with the
originals and with each other.  The theoretically expected result, net
reduction of distortion and conditional error probabilities to second order,
was indeed observed, despite imperfect coding operations which increased the
error probabilities by approximately 5\%.  Systematic study of the deviations
from the ideal behavior provided quantitative measures of different sources of
error, and good agreement was found with a numerical model.  Theoretical
questions in quantum error correction in bulk nuclear spin systems including
fidelity measures, signal strength and syndrome measurements are discussed.
\end{abstract}


% \begin{multicols}{2}[]

%%%%%%%%%%%%%%%%%%%%%%%%%%%%%%%%%%%%%%%%%%%%%%%%%%%%%%%%%%%%%%%%%%%%%%%%%%%%%
\section{Introduction}

Recent progress in experimental implementation of quantum algorithms has
demonstrated in principle that quantum computers could solve specific problems
in fewer steps than any classical
machine~\cite{Chuang98a,Chuang98b,Jones98a,Jones98b,Jones98c}.  These first
generation quantum computers were 2-spin molecules in solution.  They were
initialized, manipulated and measured at room-temperature using bulk nuclear
magnetic resonance (NMR) spectroscopy
techniques~\cite{Gershenfeld97,Cory97x,Cory97b,Chuang97e}.  Classical
redundancy in the ensemble and the discrete nature of the answers ensured that
the correct answers were obtained despite gate imperfections and moderate
rates of decoherence.  However, the accumulation of errors would be
detrimental in larger quantum computers and for longer computations in the
future, and methods to protect quantum information will be needed.

Classically, errors (bit flips) are detected and fixed by error
correcting codes.
Information is encoded redundantly, and the output contains information on
both the encoded input and the errors that have occurred, so that the errors
can be reversed.  Generalization to quantum information is non-trivial since
it is impossible to clone an arbitrary quantum bit (qubit) and to measure
quantum states without disturbing the system.  Furthermore, there is a
continuum of possible errors, and finally, entanglement can cause errors to
propagate rapidly throughout the system.

Despite the apparent difficulties, quantum error correction was shown to be
possible theoretically, and can be useful for reliable computation even when
coding operations are imperfect.  Shor and Steane~\cite{Shor95b,Steane96}
realized a major breakthrough by constructing the first quantum error
correcting codes.  Prudent use of quantum entanglement enables the information
on the errors to be obtained by non-demolition measurements without disturbing
the encoded inputs; it also enables digitization of errors.  These schemes
correct for storage errors but not for the extra errors introduced by the
coding operations.
The extension to handle gate errors and to achieve reliable computation 
with a certain accuracy threshold were subsequently developed by many
others~\cite{Shor96a,Aharonov96,Preskill97a,Kitaev97,Knill96b,Gottesman98}.

In this paper, we report experimental progress toward this elusive goal of
continued quantum computation.  We implement a simple phase error detection
scheme~\cite{Chuang95c} that {\em encodes one qubit in two} and detects a
single phase error in either one of the two qubits.  The output state is
rejected if an error is detected so that the probability to accept an
erroneous state is reduced to the smaller probability of having multiple
errors.  
Our aim is to study the effectiveness of quantum error correction in a real
experimental system, focusing on effects arising from imperfections of the
logic gates.
Therefore, the experiment is designed to eliminate potential artificial
origins of bias in the following ways.
First, we compare output states stored with and without coding (the
latter is unprotected but less affected by gate imperfections).
Second, by ensuring that all qubits used in the code decohere at nearly
the same rate, we eliminate apparent improvements brought by having an ancilla
with lifetime much longer than the original unencoded qubit.  
Third, our experiment utilizes only {\em naturally occuring} error processes. 
% 
% and thus provides realistic insight into the effectiveness of the code. 
% because phase flip errors are inevitably accompanied by other important error
% processes (such as amplitude damping) in practice.
% 
% Finally, the main error processes and their relative importance to the
% experiment are thoroughly studied and simulated to secure confidence in our
% conclusions.  
% 
Finally, the main error processes and their relative importance to the
experiment are thoroughly studied and simulated to substantiate any 
conclusions.  
In these aspects, our study differs significantly from previous
work~\cite{Cory98} demonstrating quantum error correction working only in
principle.

Using nuclear spins as qubits, we performed two sets of experiments: (1) The
``coding experiments'' in which input states were encoded, stored and decoded,
and (2) the ``control experiments'' in which encoding and decoding were
omitted.  Comparing the output states obtained from the coding and the control
experiments, both error correction by coding and extra errors caused by the
imperfect coding operations were taken into account when evaluating the actual
advantage of coding.  In our experiments, coding reduced the net error
probabilities to second order as predicted, but at the cost of small
additional errors which decreased with the original error probabilities.  
We identified the major imperfection in the logic gates to be the
inhomogeneity in the radio frequency (RF) field used for single spin
rotations.  Simulation results including both phase damping and RF field
inhomogeneity confirmed that the additional errors were mostly caused by RF
field inhomogeneity.  The causes and effects of other deviations from theory
were also studied.
{\em Tomography experiments} giving the full density matrices at major stages
of the experiments further confirmed the agreement between theoretical
expectations and the actual results.

The rest of the paper is structured into five sections: 
Section~\ref{sec:theory} consists of a comprehensive review of the background
material for the subsequent sections of the paper.  This background material
includes the phase damping model, the two bit coding scheme, and the theory of
bulk NMR quantum computation.  These are reviewed in Sections~\ref{sec:pd},
\ref{sec:2bitcode} and \ref{sec:NMR}.  Readers who are familiar with these 
subjects can skip the appropriate parts of the review.  
Section~\ref{sec:2bitNMR} describes the methods to implement the two bit 
coding scheme in NMR, and the fidelity measures to evaluate the scheme. 
Section~\ref{sec:experiment} presents the experimental details.
Section~\ref{sec:result} consists of the experimental results together with 
a thorough analysis.  The effects of coding, gate imperfections and 
the causes and effects of other discrepancies are studied in detail. 
In Section~\ref{sec:conclusion}, we conclude with a summary of our results. 
We also discuss syndrome measurement in bulk NMR, the equivalence between
logical labeling and coding, the applicability of the two bit detection
code as a correction code exploiting classical redundancy in the bulk sample 
and the signal strength issue in error correction in bulk NMR.
Sections~\ref{sec:2bitNMR} and \ref{sec:result} contain the main results of
the paper; the remainder is included for the sake of completeness and to
develop notation and terminology we believe will be helpful to the general
reader.

% \end{multicols}{2}[]
% \begin{twocolumn}
% \end{onecolumn}

%%%%%%%%%%%%%%%%%%%%%%%%%%%%%%%%%%%%%%%%%%%%%%%%%%%%%%%%%%%%%%%%%%%%%%%%%%%
\section{Theory} 
\label{sec:theory}

%%%%%%%%%%%%%%%%%%%%%%%%%%%%%%%%%%%%%%%%%%%%%%%%%%%%%%%%%%%%%%%%%%%%%%%%%%%
\subsection{Phase damping}
\label{sec:pd}

Phase damping is a decoherence process that results in the loss of coherence
between different basis states.  It can be caused by random phase shifts of
the system due to its interaction with the environment.  For example, let
$|\psi \rangle = a |0 \rangle + b |1 \rangle $ be an arbitrary pure initial
state.  A phase shift, $P$, can be represented as a rotation about the
$\hat{z}$ axis by some angle $\theta$,
\begin{equation}
	P = \mbox{ exp} \left[-\frac{i \theta}{2} \sigma_z \right] 
	= \left[ \begin{array}{cc}
	{e^{-i \theta /2}}&{0}\\{0}&{e^{i \theta /2}} 
	\end{array} \right] 
\label{eq:phasekick}
\,, 
\end{equation} 
where $\sigma_z$ is a Pauli matrix.  The resulting state is given by $P |\psi
\rangle = a e^{-i \theta /2} |0 \rangle + b e^{i \theta /2} |1 \rangle $.
Let $\rho$ be the density matrix of the initial qubit, 
\begin{equation}
	\rho = |\psi \rangle \langle \psi| 
	= \left[ \begin{array}{cc}
	{|a|^2}&{a b^*}\\{a^* b}&{|b|^2} 
	\end{array} \right] 
\label{eq:denmat}
\,. 
\end{equation} 
After the phase shift given by Eq.(\ref{eq:phasekick}), the density matrix
becomes
\begin{equation}
	\rho' = P |\psi \rangle \langle \psi| P^{\dagger} = P \rho P^{\dagger} 
	= \left[ \begin{array}{cc}
	{|a|^2}			&	{a b^*  e^{-i \theta}}
\\	{a^* b e^{i \theta}}	&	{|b|^2} 
	\end{array} \right] 
\,. 
\end{equation}

Here, we model phase randomization as a stochastic Markov process with
$\theta$ drawn from a normal distribution.  The density matrix resulting from
averaging over $\theta$ is
% {\small 
\begin{eqnarray}
	 \langle \rho' \rangle _{\theta} 
	= \int \frac{1}{\sqrt{2 \pi}s}  
	e^{-\frac{\theta^2}{2s^2}} P \rho P^{\dagger} d\theta 
	= \left[ \begin{array}{cc}
	{|a|^2}				& {a b^*  e^{-\frac{s^2}{2}}}
\\	{a^* b e^{-\frac{s^2}{2}}}	& {|b|^2} 
	\end{array} \right], 
\label{eq:phasedamp}
\end{eqnarray} 
% }
where $s^2$ is the variance of the distribution of $\theta$.  
By Markovity, the total phase shift during a time period $t$ is a random walk
process with variance proportional to $t$.
Therefore, we replace $s^2/2$ by $\lambda t$ in Eq.(\ref{eq:phasedamp}) when
the time elapsed is $t$.  
Since the diagonal and the off-diagonal elements
represent the populations of the basis states and the quantum coherence
between them, the exponential decay of the off-diagonal elements caused by
phase damping signifies the loss of coherence without net change of quanta.

Phase damping affects a mixed initial state similarly: 
\begin{equation}
	\left[ \begin{array}{cc}
	{a}	& {b^{*}}
\\	{b}	& {c} 
	\end{array} \right] ~~~ \rightarrow ~~~ 
	\left[ \begin{array}{cc}
	{a}			& {e^{-\lambda t} b^{*}}
\\	{e^{-\lambda t} b}	& {c} 
	\end{array} \right] 
\label{eq:phasedampmix}
\,,
\end{equation} 
since the density matrix of a mixed state is a weighted average of the 
constituent pure states. 

\begin{figure}[ht]
\begin{center}
\mbox{\psfig{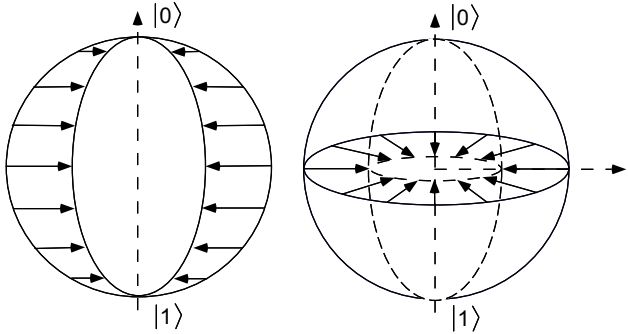}}
\vspace*{2ex}
\caption{Trajectories of different points on the Bloch sphere under the effect
	of phase damping.  Points move along perpendiculars to the $\hat
	z$-axis at rates proportional to the distances to the
	$\hat{z}$-axis.  As a result, the Bloch sphere turns into an
	ellipsoid.}
\label{fig:phasedamping}
\end{center}
\end{figure}

One can represent an arbitrary density matrix for one qubit as a Bloch vector 
$(x,y,z)$, defined to be the real coefficients in the 
{\em Pauli matrix decomposition}
\begin{equation}
	\rho=\frac{1}{2} (I + x \sigma_x + y \sigma_y + z \sigma_z) 
\label{eq:blochvec}
\end{equation} 
where the Pauli matrices are given by
\begin{equation}
	\sigma_x = \left[ \begin{array}{cc} 
	{0}&{1}\\{1}&{0} \end{array} \right],~  
	\sigma_y = \left[ \begin{array}{cc} 
	{0}&{-i}\\{i}&{0} \end{array} \right],~  
	\sigma_z = \left[ \begin{array}{cc} 
	{1}&{0}\\{0}&{-1} \end{array} \right]  
\,. 
\label{eq:pauli}
\end{equation}
The space of all possible Bloch vectors is the unit sphere known as the Bloch
sphere.  Phase damping describes the axisymmetric exponential decay of the
$\hat x$ and $\hat y$ components of any Bloch vector, as depicted in
Fig.~\ref{fig:phasedamping}.

% FIG 1 

In contrast to the above picture of phase damping as a continuous process, we
now describe an alternative model of phase damping as a discrete process.
This will facilitate understanding of quantum error correction.  The essence
is that any quantum process $\rho \rightarrow {\cal E}(\rho)$ can be written
in the {\em operator sum representation} as~\cite{Kraus70,Schumacher96a}
\begin{equation}
	{\cal E}(\rho) = \sum_k A_k \rho A_k^{\dagger}
\label{eq:osr}
\end{equation}
where the sum is over a {\em finite} number of {\em discrete} events, 
$A_k$, which are analogous to quantum jumps, and $\sum_k A_k^{\dagger} A_k$
is positive.     
For instance, Eq.(\ref{eq:phasedampmix}) describing phase damping can be
rewritten as
\begin{equation}
   	{\cal E}(\rho) = (1-p) ~ I \rho \hspace*{0.3ex} I^\dagger  
	+ ~p~ \sigma_z \hspace*{0.1ex} \rho \hspace*{0.3ex} \sigma_z^\dagger
\label{eq:opsumrep}
\,,
\end{equation}
where $p = (1-e^{-\lambda t})/2$.  
In Eq.(\ref{eq:opsumrep}), the output ${\cal E}(\rho)$ can be considered as a
(1-$p$):$p$ mixture of $\rho$ and $\sigma_z \rho \hspace*{0.3ex}
\sigma_z^{\dagger}$; in other words, ${\cal E}(\rho)$ is a mixture of 
the states after the event ``no jump'' ($I$) or ``a phase error'' ($\sigma_z$)
has occurred.  
The weights $1-p$ and $p$ are the probabilities of the two possible events.
In general, each term $A_k \rho A_k^{\dagger}$ in Eq.(\ref{eq:osr})
represents the resulting state after the event $A_k$ has occurred, with
probability $\mbox{tr}(A_k \rho A_k^{\dagger})$.  This important
interpretation is used throughout the paper.  Note that the
decomposition of ${\cal E}(\rho)$ is a mathematical interpretation rather
than a physical process.
The component states $A_k \rho A_k^{\dagger}$ of ${\cal E}(\rho)$ are not
generally obtainable because they are not necessarily orthogonal to each
other. 

We emphasize that Eqs.(\ref{eq:phasedampmix}) and (\ref{eq:opsumrep}) describe
the same physical process.  Eq.(\ref{eq:opsumrep}) provides a discrete
interpretation of phase damping, with the continuously changing parameter
$e^{-\lambda t}$ embedded in the probabilities of the possible events.

For a system of multiple qubits, we {\em assume} independent
decoherence on each qubit.  For example, for two qubits $A$ and $B$
with error probabilities $p_a$ and $p_b$, the joint process is given
by
\begin{eqnarray}
 	{\cal E}(\rho) & = & (1-p_a)(1-p_b) ~(I \otimes I)~\rho~(I \otimes I) 
\nonumber
\\
	& + &  (1-p_a) ~ p_b ~~~(I \otimes \sigma_z)~\rho~(I \otimes \sigma_z)
\nonumber
\\
	& + &  p_a ~ (1-p_b) ~~~(\sigma_z \otimes I)~\rho~(\sigma_z \otimes I)
\nonumber
\\
	& + &  p_a ~ p_b ~~~~~ (\sigma_z \otimes \sigma_z) ~\rho~ (\sigma_z 
				\otimes \sigma_z)
\,,
\label{eq:opsumrep2}
\end{eqnarray}
where $\rho$ now denotes the $4$ $\times$ $4$ density matrix for the two
qubits.  The events $\sigma_z \otimes I$ and $I \otimes \sigma_z$ are first
order errors, while $\sigma_z \otimes \sigma_z$ is second order.  First and
second order events occur with probabilities linear and quadratic in the small
error probabilities.

Having described the noise process, we now proceed to describe a coding scheme
that will correct for it.

%%%%%%%%%%%%%%%%%%%%%%%%%%%%%%%%%%%%%%%%%%%%%%%%%%%%%%%%%%%%%%%%%%%%%%%%%%%%%
\subsection{The two bit phase damping detection code} 
\label{sec:2bitcode}

Quantum error correction is similar to its classical analogue in many
aspects.  First, the input is encoded in a larger system which goes through
the decoherence process, such as transmission through a noisy channel or
storage in a noisy environment.  The encoded states (codewords) are chosen
such that information on the undesired changes (error syndromes) can be
obtained in the extra degrees of freedom in the system upon decoding.  Then
corrections can be made accordingly.  However, in contrast to the classical
case, quantum errors occur in many different forms such as phase flips -- and
not just as bit flips.  Furthermore, the quantum information must be
preserved without ever measuring it, because measurement which obtains
information about a quantum state inevitably disturbs it.  There are
excellent references on the theory of quantum error
correction~\cite{Ekert96b,Knill96,Bennett96a,Nielsen96c,Gottesman97}.  
We limit the present discussion to detection codes only.

For a code to detect errors, it suffices to choose the codeword space ${\cal
C}$ such that all errors to be detected map ${\cal C}$ to its orthogonal
complement.  In this way, detection can be done
unambiguously by a projection onto ${\cal C}$ {\em without} distinguishing
individual codewords; hence without disturbing the encoded information.  
To make this concrete, consider the code~\cite{Chuang95c}
\begin{eqnarray}
	|0_L \rangle & = &\frac{1}{\sqrt{2}} (|00 \rangle +|11 \rangle )
\\
	|1_L \rangle & = &\frac{1}{\sqrt{2}} (|01 \rangle +|10 \rangle )
\,, 
\label{eq:code}
\end{eqnarray}
where the subscript $L$ denotes logical states.  An arbitrary encoded 
qubit is given by 
\begin{eqnarray}
	|\psi \rangle 
	& = & a |0_L \rangle + b |1_L \rangle \\
	& = & \frac{1}{\sqrt{2}} \left[\rule{0pt}{2.4ex} 
	a (|00 \rangle +|11 \rangle ) + b (|01 \rangle +|10 \rangle ) \right]
\label{eq:psi}
\,.
\end{eqnarray}
After the four possible errors in Eq.(\ref{eq:opsumrep2}), the possible 
outcomes are
{\small
\begin{eqnarray}
	|\psi_{II} \rangle = & I \otimes I~ |\psi \rangle  = & 
	a~\frac{ |00 \rangle + |11 \rangle }{\sqrt{2}} 
	+ b~\frac{ |01 \rangle +|10 \rangle }{\sqrt{2}}
\label{eq:iipsi} \\ 
	|\psi_{ZI} \rangle = & \sigma_z \otimes I~ |\psi \rangle = & 
	a~\frac{|00 \rangle -|11 \rangle }{\sqrt{2}} 
	+ b~\frac{|01 \rangle -|10 \rangle }{\sqrt{2}}
\label{eq:zipsi} \\
	|\psi_{IZ} \rangle = & I \otimes \sigma_z |\psi \rangle = & 
	a~\frac{|00 \rangle -|11 \rangle }{\sqrt{2}} 
	+ b~\frac{ -|01 \rangle +|10 \rangle }{\sqrt{2}} 
\label{eq:izpsi} \\
	|\psi_{ZZ} \rangle = & \sigma_z \otimes \sigma_z |\psi \rangle = & 
	a~\frac{|00 \rangle +|11 \rangle }{\sqrt{2}} 
	+ b~\frac{-|01 \rangle -|10 \rangle }{\sqrt{2}}
\label{eq:zzpsi}
\,,
\end{eqnarray}
}
with the {\em first order} erroneous states $|\psi_{ZI} \rangle $ and
$|\psi_{IZ} \rangle $ orthogonal to the correct state $|\psi_{II} \rangle $.
Therefore, it is possible to distinguish (\ref{eq:iipsi}) from
(\ref{eq:zipsi}) and (\ref{eq:izpsi}) by a projective measurement during
decoding, which is described next.

\begin{figure}[ht]
\begin{center}
\mbox{\psfig{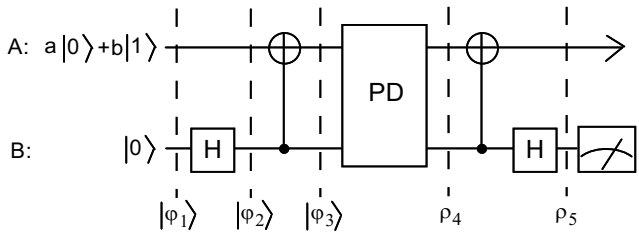}}
\end{center}
\caption{Circuit for encoding and decoding.  Qubit $A$ is the input qubit.  
	$H$ is the Hadamard transformation, and the symbol next to $H$ is a
	controlled-not with the dot and circle being the control and target
	bits.  $|\psi_{1-3} \rangle $ are given by
	Eqs.(\ref{eq:psi1})-(\ref{eq:psi3}). $\rho_4$, $\rho_5$ are mixtures
	of the states in Eqs.(\ref{eq:iipsi})-(\ref{eq:zzpsi}) and in
	Eqs.(\ref{eq:phi})-(\ref{eq:zzphi}).  A phase error in either one of
	the qubits will be revealed by qubit $B$ being in $|1 \rangle $ after
	decoding, and in that case, qubit $A$ will be rejected.  }
\label{fig:codingcircuit}
\end{figure}

The encoding and decoding can be performed as follows.  We start with an
arbitrary input state and a ground state ancilla, represented as qubits $A$
and $B$ in the circuit in Fig.~\ref{fig:codingcircuit}.

% FIG 2

To encode the input qubit, a Hadamard transformation, $H$, is applied 
to the ancilla, followed by a controlled-not from the ancilla to qubit $A$ 
(written as $C\!\!N_{ba}$).  Let $A$ and $B$ be the first and second label. 
Then, the two operations have matrix representations 
\begin{equation}
	H = \frac{1}{\sqrt{2}} \left[ \begin{array}{cc} 
	{1}&{1}\\{1}&{-1} \end{array} \right]~~  
	C\!\!N_{ba} = \left[ \begin{array}{cccc} 
	{1}&{0}&{0}&{0}\\{0}&{0}&{0}&{1}\\
	{0}&{0}&{1}&{0}\\{0}&{1}&{0}&{0} \end{array} \right] 
\,, 
\end{equation}
and the qubits transform as (Fig.~\ref{fig:codingcircuit})
\begin{eqnarray}
	|\psi_1 \rangle & = & (a |0 \rangle + b |1 \rangle )|0 \rangle 
\label{eq:psi1}
\\	\stackrel{I \otimes H}{\longrightarrow} |\psi_2 \rangle 
	& = & \frac{1}{\sqrt{2}}(a |0 \rangle + b |1 \rangle )
	(|0 \rangle +|1 \rangle )
\label{eq:psi2}
\\	\stackrel{C\!\!N_{ba}}{\longrightarrow} |\psi_3 \rangle 
	& = &\frac{1}{\sqrt{2}}(a |0 \rangle + b |1 \rangle )|0 \rangle 
	+ (a |1 \rangle + b |0 \rangle )|1 \rangle  
\label{eq:psi3}
\\	& =  & \frac{1}{\sqrt{2}}(a (|00 \rangle + |11 \rangle ) 
	+ b (|01 \rangle + |10 \rangle )
\label{eq:psi33}
\,, 
\end{eqnarray}
where Eq.(\ref{eq:psi33}) is the desired encoded state.

The decoding operation is the inverse of the encoding operation (see
Fig.~\ref{fig:codingcircuit}) so as to recover the input $(a |0 \rangle + b
|1 \rangle )|0 \rangle $ in the absence of errors.  Phase errors lead to other 
decoded outputs.  The possible decoded states are given by:
\begin{eqnarray}
	|\psi_{II} \rangle & \stackrel{dec}{\Rightarrow} 
				& (a |0 \rangle + b |1 \rangle ) |0 \rangle 
\label{eq:phi} 
\\
	|\psi_{ZI} \rangle &  \Rightarrow  
				&(a |0 \rangle - b |1 \rangle ) |1 \rangle 
\label{eq:ziphi} 
\\
	|\psi_{IZ} \rangle &  \Rightarrow 
				&(a |0 \rangle + b |1 \rangle ) |1 \rangle 
\label{eq:izphi} 
\\
	|\psi_{ZZ} \rangle &  \Rightarrow 
				&(a |0 \rangle - b |1 \rangle ) |0 \rangle 
\label{eq:zzphi}
\,.
\end{eqnarray}

Note that the ancilla becomes $|1 \rangle $ upon decoding if and only if a
{\em single} phase error has occurred.  Moreover, qubits A and B are in
product states but they are classically correlated.  Therefore, syndrome can
be read out by a projective measurement on $B$ without measuring the encoded
state.  The decoding operation transforms the codeword space and its
orthogonal complement to the subspaces spanned by $|0 \rangle $ and $|1
\rangle $ in qubit $B$, while all the encoded information, either with or
without error, goes to qubit $A$.

We illustrate the role of entanglement in the digitization and detection of
errors as follows.  Suppose the error is a phase shift on qubit $A$: $|0
\rangle \rightarrow |0 \rangle $, $|1 \rangle \rightarrow e^{i \theta}|1
\rangle $.  Then, the encoded state becomes
\begin{eqnarray}
	& & \frac{1}{\sqrt{2}} \left[\rule{0pt}{2.4ex} 
	  a (|00 \rangle + e^{i \theta} |11 \rangle ) 
	+ b (e^{i \theta} |10 \rangle + |01 \rangle ) \right]
\\
	& = & \frac{1}{\sqrt{2}}\frac{1+e^{i \theta}}{2} 
	\left[\rule{0pt}{2.4ex} a (|00 \rangle +|11 \rangle ) 
	+ b (|10 \rangle + |01 \rangle ) \right]  
\nonumber
\\
 	& + &  \frac{1}{\sqrt{2}}\frac{1-e^{i \theta}}{2} 
	\left[\rule{0pt}{2.4ex} a (|00 \rangle -|11 \rangle ) 
	+ b (-|10 \rangle + |01 \rangle ) \right]
\,.
\end{eqnarray}
The decoded state is now a superposition of the states given by  
Eqs.(\ref{eq:phi}) and (\ref{eq:ziphi}):
\begin{equation}
	\frac{1}{\sqrt{2}}\frac{1+e^{i \theta}}{2}
	(a |0 \rangle + b |1 \rangle ) |0 \rangle + 
	\frac{1}{\sqrt{2}}\frac{1-e^{i \theta}}{2} 
	(a |0 \rangle - b |1 \rangle ) |1 \rangle 
\nonumber
\,. 
\end{equation}
Measurement of qubit $B$ projects it to either $|0 \rangle$ or $|1
\rangle$. {\em Because of entanglement}, qubit $A$ is projected to having no
phase error, or a complete phase flip.

We quantify the error correcting effect of coding using the discrete
interpretation of the noise process, leaving a full discussion of the fidelity
to Section~\ref{sec:2bitNMR}.  
Recall from Eq.(\ref{eq:opsumrep2}) that the errors $I \otimes I$, $I \otimes
\sigma_z$, $\sigma_z \otimes I$, and $\sigma_z \otimes \sigma_z$ occur with
probabilities $(1-p_a)(1-p_b)$, $(1-p_a) p_b$, $p_a (1-p_b)$ and $p_a p_b$
respectively, and only in the first and the last cases will the output state
be accepted.  The probability of accepting the output state is $(1-p_a)(1-p_b)
+ p_a p_b$ whereas the probability of accepting the correct state is
$(1-p_a)(1-p_b)$.  The conditional probability of a correct, accepted state is
therefore 
\begin{equation} 
	\frac{(1-p_a)(1-p_b)}{(1-p_a)(1-p_b) + p_a p_b} \approx 1 - p_a p_b 
\, 
\end{equation} 
for small $p_a$, $p_b$.  The code improves the conditional error probability
to second order, as a result of screening out the first order erroneous
states.

We conclude this section with a discussion of some properties of the two bit
code.
First, the code also applies to mixed input states since the
code preserves all constituent pure states in the mixed input.
Second, we show here that two qubits are the minimum required to encode one
qubit and to detect any phase errors.  Let ${\cal C}$ be the 2-dimensional
codeword space and $E$ be a non-trivial error to be detected.  For phase
damping, $E$ is unitary and therefore $E {\cal C}$ is also 2-dimensional.
Moreover, ${\cal C}$ and $E {\cal C}$ must be orthogonal if $E$ is to be
detected.  Therefore the minimum dimension of the system is $4$, which
requires two qubits.
However, using only two qubits implies other intrinsic limitations.  First,
the code can detect but cannot distinguish errors.  Therefore, it cannot
correct errors.  Moreover, $|\psi_{IZ} \rangle $ decodes to a correct state in 
spin $A$ which is rejected.  These affect the
absolute fidelity (the overall probability of successful recovery) but not the
conditional fidelity (the probability of successful recovery if the state is
accepted).  Second, the error $\sigma_z \otimes \sigma_z$ cannot be detected.
This affects both fidelities but only in second order.  To understand why
these limitations are intrinsic, let $\{E_k\}$ be the set of non-trivial
errors to be detected.  Since $E_k {\cal C}$ has to be orthogonal to ${\cal
C}$ for all $k$, and since ${\cal C}$ has a unique orthogonal complement of
dimension 2 in a two bit code, it follows that all $E_k {\cal C}$ are equal,
and it is impossible to distinguish (and correct) the different errors.  By
the same token, for any distinct errors $E_{k'}$ and $E_k$, $E_{k'} E_k {\cal
C} = {\cal C}$ because they are both orthogonal to $E_{k'} {\cal C}$, which
has a unique 2-dimensional orthogonal complement.  Therefore, a two bit code
that detects single phase errors can never detect double errors.
Finally, since a detection code cannot correct errors, it can only improve the
conditional fidelity of the {\em accepted} states but not the absolute
fidelity.  Here, we only remark that the conditional fidelity is a better
measure in our experiments due to the bulk system used to implement the two
bit code.  A discussion of fidelity measures in our experiments and quantum
error correction in bulk systems will be postponed until
Sections~\ref{sec:2bitNMR} and \ref{sec:conclusion}.  The system in
which the two bit code is implemented will be described next.

%%%%%%%%%%%%%%%%%%%%%%%%%%%%%%%%%%%%%%%%%%%%%%%%%%%%%%%%%%%%%%%%%%%%%%%%%%%
\subsection{Bulk NMR Quantum Computation}
\label{sec:NMR}

Nuclear spin systems are good candidates for quantum computers for many
reasons.  Nuclear spins can have long coherence times.  Coupled operations
involving multiple qubits are built in as coupling of spins within molecules.
Complex sequences of operations can be programmed and carried out easily using
modern spectrometers.  However, the signal from a single spin is so weak that
detection is not feasible with current technology unless a bulk sample of
identical spin systems is measured.  These identical systems run the same
quantum computation in classical parallelism.  Computation can be performed at
room temperature starting with mixed initial states by distilling the signal
of the small excess population in the desired ground state.  How NMR quantum
computation can be done is described in detail in the following.

\begin{figure}[ht]
\centerline{\mbox{\psfig{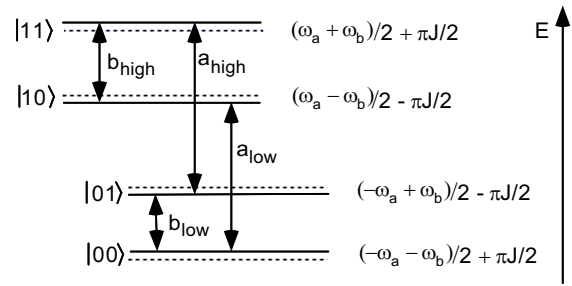}}}
\vspace*{1.5ex}
\caption{Energy diagram for the two-spin nuclear system.  The transitions
	labelled $a_{low}$, $a_{high}$, $b_{low}$ and $b_{high}$ refer to
	transitions $(|0 \rangle \leftrightarrow |1 \rangle )|0 \rangle $,
	$(|0 \rangle \leftrightarrow |1 \rangle )|1 \rangle $, $|0 \rangle (|0
	\rangle \leftrightarrow |1 \rangle )$ and $|1 \rangle (|0 \rangle
	\leftrightarrow |1 \rangle )$ respectively.}
\label{fig:energydiagram}
\end{figure}

\hfill \\
\noindent {\bf The quantum system (hardware)}~~
In our two bit NMR system, $|0 \rangle $ and $|1 \rangle $ describe the ground
and excited states of the nuclear spin (the states aligned with and against an
externally applied static magnetic field ${\bf B}_0$ in the $+ \hat z$
direction).  As in the previous section, we call the spins denoted by the
first and second registers $A$ and $B$.  The {\em reduced} Hamiltonian for our
system is well approximated by ($\hbar=1$)~\cite{Slichter,Abragam} (see also
Fig.~\ref{fig:energydiagram})
\begin{equation}
	{\cal H} = - \frac{\omega_a}{2} {\sigma}_{z} \otimes I 
 			- \frac{\omega_b}{2} I \otimes {\sigma}_{z} 
 			+ \frac{\pi J}{2} {\sigma}_{z} \otimes {\sigma}_{z} 
 			+ {\cal H}_{env} 
\,.
\label{eq:nmrhamiltonian}
\end{equation}
The first two terms on the right hand side of Eq.(\ref{eq:nmrhamiltonian}) are
Zeeman splitting terms describing the free precession of spins $A$ and $B$
about the $-\hat z$ direction with frequencies $\omega_a/2\pi$ and
$\omega_b/2\pi$.  The third term describes a spin-spin coupling of $J$ Hz
which is electron mediated.  It is known as the $J$ coupling.  ${\cal
H}_{env}$ represents coupling to the reservoir, such as interactions with
other nuclei, and higher order terms in the spin-spin coupling.

\hfill \\
\noindent {\bf Universal set of quantum logic gates}~~ 
A set of logic gates is {\em universal} if any operation can be approximated 
by some suitable sequence of gates chosen from the set.  
Depending on whether computation is fault-tolerant or not, the minimun 
requirements for universality are different.  
In the latter case, any coupled two-qubit operation together with the
set of all single qubit transformations form a {\em universal set} of
quantum gates~\cite{Div95,Barenco95a,Deutsch95,Barenco95b}.
Both requirements are satisfied in NMR as follows.

\hfill \\
\noindent {\bf Single qubit rotations}~~
Spin flip transitions between the two energy eigenstates can be induced by
pulsed radio frequency (RF) magnetic fields.  These fields, oriented in the
$\hat x \hat y$-plane perpendicular to ${\bf B}_0$, selectively address either
$A$ or $B$ by oscillating at angular frequencies $\omega_a$ or $\omega_b$.  In
the classical picture, an RF pulse along the axis $\hat \eta$ rotates a spin
about $\hat \eta$ by an angle $\theta$ proportional to the product of the
pulse duration and amplitude of the oscillating magnetic field.  In the
quantum picture, the rotation operator $e^{-i \frac{\theta}{2} \vec{\sigma}
\cdot \hat{\eta}}$ {\em rotates} the Bloch vector (Eq.(\ref{eq:blochvec}))
likewise.  Throughout the paper, we denote rotations of $\pi/2$ along the
$\hat{x}$ and $\hat{y}$ axes for spins $A$ and $B$ by $X_a$, $Y_a$, $X_b$, and
$Y_b$ with respective matrix representations $e^{-i \frac{\pi}{4} \sigma_x
\otimes I}$, $e^{-i \frac{\pi}{4} \sigma_y \otimes I}$, $e^{-i \frac{\pi}{4} I
\otimes \sigma_x}$, and $e^{-i \frac{\pi}{4} I \otimes \sigma_y}$.  The
rotations in the reverse directions are denoted by an additional ``bar'' above
the symbols of the original rotations, such as $\overline{X}_a$.  The angle of
rotation is given explicitly when it differs from $\pi/2$.  This set of
rotations generates the Lie group of all single qubit operations, $SU(2)$.
For example, the Hadamard transformation can be written as $ H = i e^{-i
\frac{3\pi}{4} \sigma_y} e^{i \frac{\pi}{2} \sigma_x}$ which can be
implemented in two pulses.

\hfill \\
\noindent {\bf Coupled operations}~~
Quantum entanglement, essential to quantum information processing, can be 
naturally created by the time evolution of the system.  In the
respective rotating frames of the spins (tracing the free precession
of the uncoupled spins), only the $J$-coupling term, $e^{-i \frac{\pi J
t}{2} \sigma_z \otimes \sigma_z}$, is relevant in the time evolution.
Entanglement is created because the evolution depends on the
state of {\em both} spins.  A frequently used coupled ``operation''
is a time delay of $\frac{1}{2J}$, denoted by $\tau$, which
corresponds to the evolution $e^{-i \frac{\pi}{4}
\sigma_z \otimes \sigma_z}$.  For instance, appending $\tau$ with
the {\em single qubit rotations} $e^{i \frac{\pi}{4} \sigma_z \otimes I}$ and
$e^{i \frac{\pi}{4} I \otimes \sigma_z}$ about the $-\hat{z}$ axes of 
$A$ and $B$, we implement the unitary operation
\begin{equation}
	\chi = e^{i \pi/4} \left[ \begin{array}{cccc} 
	{1}&{0}&{0}&{0}\\{0}&{1}&{0}&{0}\\
	{0}&{0}&{1}&{0}\\{0}&{0}&{0}&{-1} \end{array} \right] 
\,,
\end{equation}
which is a cross-phase modulation between the two qubits. 
Together with the set of all single qubit transformations, 
$\chi$ completes the requirement for universality. 
For instance, the controlled-not, $C\!\!N_{ba}$, mentioned in the 
previous section, can be written as $(H \otimes I)~ \chi~ (H \otimes I)$, 
and can be implemented by concatenating the sequences for each constituent 
operation.  
It is also crucial that the free evolution which leads to creation of
entanglement can be reversed by applying 
{\em refocusing} $\pi$-pulses, such that creation of entanglement
between qubits can be stopped.  This technique will be described in detail
later.
 
\hfill \\
\noindent {\bf Measurement}~~
The measured quantity in NMR experiments is the time varying voltage
induced in a pick-up coil in the $\hat x \hat y$-plane:
{\small 
\begin{eqnarray}
	V(t) = - V_0 Tr \left[\rule{0pt}{2.1ex}\right. & e^{-i{\cal H}t}
		\rho(0) e^{i{\cal H}t} &
\nonumber
\\ 
	& \times (\hspace*{0.5ex}(i \sigma_{x} + \sigma_{y}) & \otimes I 
	  	   + I \otimes (i \sigma_{x} + \sigma_{y}) \hspace*{0.5ex}) 
			\left.\rule{0pt}{2.1ex}\right] .
\label{eq:FID} 
\end{eqnarray}
}
The signal $V(t)$, known as the {\em free induction decay} (FID), is recorded
with a phase-sensitive detector.  In Eq.(\ref{eq:FID}), the onset of
acquisition of the FID is taken to be $t=0$.  If the density matrix $\rho(0)$
has Pauli matrix decomposition $\rho(0) = \sum_{i,j=0}^3 c_{ij} \sigma_i
\otimes \sigma_j$ where $\sigma_{0,1,2,3}$ are the identity matrix and
$\sigma_{x,y,z}$ respectively, then the spectrum of $V(t)$ has four lines at
frequencies $\frac{\omega_a}{2 \pi}+\frac{J}{2}$, $\frac{\omega_a}{2
\pi}-\frac{J}{2}$, $\frac{\omega_b}{2 \pi}+\frac{J}{2}$, $\frac{\omega_b}{2
\pi}-\frac{J}{2}$, with corresponding integrated areas (``peak integrals'')
\begin{eqnarray}
	I_{a_{high}} & = &-\left[\rule{0pt}{2.4ex} 
				i (c_{10} - c_{13}) + c_{20} - c_{23} \right]
\label{eq:ahigh}
\\
	I_{a_{low}} & = &-\left[\rule{0pt}{2.4ex} 
				i (c_{10} + c_{13}) + c_{20} + c_{23} \right]
\label{eq:alow}
\\
	I_{b_{high}} & = &-\left[\rule{0pt}{2.4ex} 
				i (c_{01} - c_{31}) + c_{02} - c_{32} \right]
\label{eq:bhigh}
\\
	I_{b_{low}} & = &-\left[\rule{0pt}{2.4ex} 
				i (c_{01} + c_{31}) + c_{02} + c_{32} \right]
\label{eq:blow}
\,.
\end{eqnarray}
Note that the expression $c_{10} - c_{13}$~($c_{10} + c_{13}$) occuring in the
{\em high} ({\em low}) frequency line of spin $A$ is the coefficient of
$\sigma_x \otimes |1 \rangle \langle 1|~(|0 \rangle \langle 0|)$ in $\rho(0)$.
Similarly, $c_{20} - c_{23}$~($c_{20} + c_{23}$) is the coefficient of
$\sigma_y \otimes |1 \rangle \langle 1|~(|0 \rangle \langle 0|)$.  They
signify transitions between the states $|0 \rangle \leftrightarrow |1 \rangle
$ for spin $A$ when spin $B$ is in $|1 \rangle $ ($|0 \rangle $).  Similar
observations hold for the high and low lines of spin $B$.

% FIG 3

\hfill\\
\noindent {\bf Thermal States}~~
In bulk NMR quantum computation at room temperature, a pure initial state is
not available due to large thermal fluctuations ($\hbar \omega_a,~\hbar
\omega_b \ll kT$).  Instead, a convenient class of initial states arises from
the thermal equilibrium states (thermal states).
In the energy eigenbasis, the density matrix is diagonal with diagonal entries
proportional to the Boltzmann factors, in other words, $\rho_{th} =
\frac{1}{{\cal Z}} e^{-\frac{{\cal H}}{kT}}$, where $kT$ is the thermal energy
and ${\cal Z}$ is the partition function normalization factor.
At room temperature, $ \langle \frac{{\cal H}}{kT} \rangle \approx 10^{-6}$, 
${\cal Z} \approx \mbox{Dim}(\rho_{th})$ and 
$\rho_{th} \approx (I - \frac{{\cal H}}{kT})/\mbox{Dim}(\rho_{th})$ 
to first order.  For most of the time, the identity term in the above
expansion is omitted in the analysis for two reasons.  First, it does not
contribute to any signal in Eqs.(\ref{eq:ahigh})-(\ref{eq:blow}).  Physically,
it represents a completely random mixture which is {\em isotropic} and, by
symmetry, has no {\em net} magnetization at any time.
Second, the identity is invariant under a wide range of processes.  Processes
which satisfy ${\cal E}(I) = I$ are called {\em unital}.  These include all
unitary transformations and phase damping.  Under unital processes, the
evolution of the state is completely determined by the evolution of the 
deviation from identity~\cite{Gershenfeld97,Chuang97e}, in which case 
the identity can be neglected.

The unitality assumption is a good approximation in our system.  The main
cause of non-unitality is amplitude damping.  In general, for
a non-unital but trace preserving process ${\cal E}$, the observable evolution
of the deviation can be understood as follows
\cite{Nielsen97}.
Rewriting 
\begin{equation}
	\rho = \upsilon I + \rho_\delta
\end{equation}
where $\rho_\delta = \rho - \upsilon I$ is the traceless deviation from 
the identity and $\upsilon = 1/\mbox{Dim}(\rho)$.  Then, 
\begin{eqnarray}
	{\cal E}(\rho) & = & \upsilon I + \tilde{\rho}_\delta
\\
	\tilde{\rho}_\delta & = & \upsilon ({\cal E}(I) - I) 
				+ {\cal E}(\rho_\delta)
\label{eq:deltaevolve}
\,.
\end{eqnarray}
The observed evolution of the deviation is $\rho_{\delta} \rightarrow
\tilde{\rho}_\delta$.  The second term in Eq.(\ref{eq:deltaevolve}) comes from
the evolution of $\rho_\delta$ when the identity is neglected, and the first
term is the correction due to non-unitality.  In our experiment, amplitude
damping is slow compared to all other time scales, therefore ${\cal E}(I) - I$
is small and can be treated as a small extra distortion of the state when
making the unitality assumption.

\hfill \\
\noindent {\bf Temporal labelling}~~ 
One convenient method to create arbitrary {\em initial deviations} from the
thermal mixture is temporal labelling~\cite{Chuang97e,Knill97}.  The idea is
to add up the results of a series of experiments that begin with different
preparation pulses before the intended experiment, so as to cancel out the
signals from the undesired components in the initial thermal mixture.
Mathematically, let $\{P_k\}$ be the set of initial pulses and ${\cal
E}(\rho)$ be the intended computation process.  By linearity, $\sum_k {\cal
E}(P_k \rho_{th} P_k^{\dagger}) = {\cal E}(\sum_k P_k \rho_{th}
P_k^{\dagger})$.  Summing over the experimental results (on the left side) is
equivalent to performing the experiment with initial state $\sum_k P_k
\rho_{th} P_k^{\dagger}$ (on the right side).  Temporal labelling assumes the
repeatability of the experiments, which is true up to small fluctuations.

\begin{figure}[ht]
\begin{flushleft}
\centerline{\mbox{\psfig{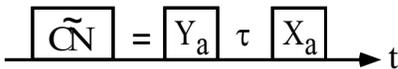}}}
\caption{Pulse sequence for $\tilde{C\!\!N}$.  Time runs from left to right.
\hfill}
\end{flushleft}
\label{fig:cnshort}
\end{figure}

\hfill \\
\noindent {\bf Example}~~
As an example of the above theories, consider applying the pulse sequence 
in Fig.~\ref{fig:cnshort} to the thermal state.

% FIG 4 

The pulses are short compared to other relevant time scales.  Therefore, other
changes of the system during the pulses are ignored.  The unitary operation
implemented by the above sequence is given by
\begin{eqnarray}
\tilde{C\!\!N} & = & e^{-i \frac{\pi}{4} \sigma_x \otimes I}
 e^{-i \frac{\pi}{4} \sigma_z \otimes \sigma_z}
 e^{-i \frac{\pi}{4} \sigma_y \otimes I} 
\\
& = & \frac{1}{\sqrt{2}} \left[ \begin{array}{cccc}

	{1-i}&{0}&{0}&{0}
\\
	{0}&{0}&{0}&{-1-i}
\\
	{0}&{0}&{1+i}&{0}
\\
 	{0}&{1-i}&{0}&{0}

	\end{array} \right]
\,,
\end{eqnarray}
similar to $C\!\!N_{ba}$ described in Section \ref{sec:NMR}. 

The deviation density matrix of the thermal state is proportional to 
$-{\cal H}$.  Neglecting the $J$-coupling term, which is much smaller 
than the Zeeman terms,    
\begin{eqnarray}
	& & \rho_{th} \sim \frac{\omega_a}{2} ~\sigma_z \otimes I 
	+ \frac{\omega_b}{2} ~I \otimes \sigma_z 
\label{eq:rhoth}
\\
	& = & \frac{1}{2} \mbox{ Diag}(\omega_a+\omega_b,
	\omega_a-\omega_b, -\omega_a+\omega_b, -\omega_a-\omega_b)
\,
\end{eqnarray}
where ``Diag'' indicates a diagonal matrix with the given elements.   
For simplicity, we omit the proportionality constant in Eq.(\ref{eq:rhoth}), 
and rename the right hand side as $\rho_{th}$.  
The sequence transforms $\rho_{th}$ to 
\begin{eqnarray}
	&  & \rho_{cn}  =  
		\tilde{C\!\!N} \rho_{th} \tilde{C\!\!N}^{\dagger}  
\\
 		& = & \frac{1}{2} \mbox{ Diag}(\omega_a+\omega_b, 
		-\omega_a-\omega_b,-\omega_a+\omega_b, \omega_a-\omega_b)
\\
		& = & \frac{\omega_a}{2} ~\sigma_z \otimes \sigma_z 
		+ \frac{\omega_b}{2} ~I \otimes \sigma_z 
\label{eq:rhocn}
\,, 
\end{eqnarray}
in which the populations of $|01 \rangle $ and $|11 \rangle $ are
interchanged.  Therefore, $\tilde{C\!\!N}$ and $C\!\!N_{ba}$ act in the
same way {\em on the thermal state}.

\begin{figure}[ht]
\centerline{\mbox{\psfig{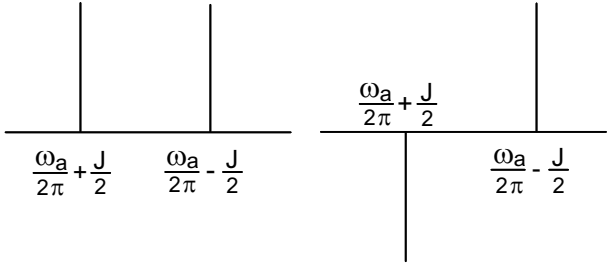}}}
\vspace*{1ex}
\caption{(a) Spectrum of $A$ after a readout pulse on the thermal state. 
	 (b) Spectrum of $A$ after $\tilde{C\!\!N}$ and a readout pulse.} 
\label{fig:cnspect}
\end{figure}

By inspection of Eqs.(\ref{eq:rhoth}) and (\ref{eq:rhocn}), it
can be seen that $\rho_{th}$ and $\rho_{cn}$ have zero 
peak integrals given by Eqs.(\ref{eq:ahigh})-(\ref{eq:blow}).  To obtain more 
information about the states, a readout pulse $X_a$ can be applied 
to transform the two states to
\begin{equation}
	\rho_{th}' = - \frac{\omega_a}{2} ~\sigma_y \otimes I 
	+ \frac{\omega_b}{2} ~I \otimes \sigma_z 
\end{equation}
and
\begin{equation} 
	\rho_{cn}' = - \frac{\omega_a}{2} ~\sigma_y \otimes \sigma_z 
		+ \frac{\omega_b}{2} ~I \otimes \sigma_z 
\,. 
\end{equation}
Now, in $\rho_{th}'$ is a term $\sigma_y \otimes I$ with coefficient
$c_{20} = - \frac{\omega_a}{2}$ which contributes to two spectral lines
at $\frac{\omega_a}{2 \pi} \pm \frac{J}{2}$ with equal and positive,
real peak integrals.  
The readout pulse transforms the unobservable coefficient $c_{30}$ in
$\rho_{th}$ to the observable $-c_{20}$ in $\rho_{th}'$, yielding 
information on {\em the state before the readout pulse}.
Similarly, $\rho_{cn}'$ has a $\sigma_y \otimes \sigma_z$ term with
coefficient $c_{23} = - \frac{\omega_a}{2}$ which gives rise to two spectral
lines with real and opposite peak integrals (Fig.~\ref{fig:cnspect}).  All
outputs in our experiments are peak integrals of this type carrying
information on the decoded states.

% FIG 5

%%%%%%%%%%%%%%%%%%%%%%%%%%%%%%%%%%%%%%%%%%%%%%%%%%%%%%%%%%%%%%%%%%%%%%%
\section{Two bit code in NMR}
\label{sec:2bitNMR}

We now describe how the two bit code experiment can be implemented in an
ensemble of two-spin systems.  Modifications of the standard theories in
Section \ref{sec:2bitcode} are needed.  These include methods for state
preparation, designing encoding and decoding pulse sequences, methods
to store the qubit with controllable phase damping, and finally methods to
read out the decoded qubit.  Fidelity measures for deviation density matrices
are also defined.
 
\begin{figure}[ht]
\centerline{\mbox{\psfig{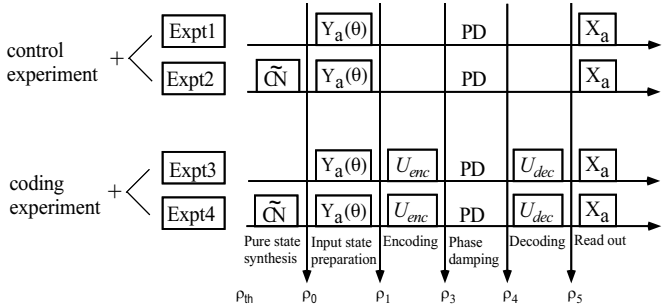}}}
\caption{Schematic diagram for the two-bit code experiment.
         $\tilde{C\!\!N}$ is used to prepare the initial state.
         $Y_a(\theta)$ is a variable angle rotation applied to prepare an
         arbitrary input state, which is then subject to phase damping
         (PD). In the coding experiment, encoding and decoding operations, 
         $U_{enc}$ and $U_{dec}$, are performed before and after phase damping,
         whereas in the control experiment, these operations are omitted.
         $X_a$ is used as a readout pulse on $A$ to determine the output
         state $\rho_5$ in spin $A$.  $\rho_i$ corresponds to $|\psi_i \rangle
         $ or $\rho_i$ in Fig.~\ref{fig:codingcircuit}.  Details are described
         in the text.}

\label{fig:scheme}
\end{figure}

Spins $A$ and $B$ are designated to be the input and the ancilla qubits
respectively.  The output states of spin $A$ are reconstructed from the peak
integrals at frequencies $\omega_a/2\pi \pm J/2$.  Fig.~\ref{fig:scheme}
schematically summarizes the major steps in the experiments, with details
given in the text.

Some notation is defined as follows.  Initial states and input states refer to
$\rho_0$ and $\rho_1$ in Fig.~\ref{fig:scheme}.  The phrase ``ideal case''
refers to the scenario of having perfect logical operations throughout the
experiments and pure phase damping during storage.

% FIG 6

\hfill \\
\noindent {\bf Initial state preparation}~~ 
It is necessary to initialize spin $B$ to $|0 \rangle $ before the experiment.
This can be done with temporal labeling using two experiments: the first
experiment starts with no additional pulses; the second experiment starts with
$\tilde{C\!\!N}$ (Fig.~\ref{fig:cnshort}).  Therefore, the equivalent initial
state is $\rho_{th}+\rho_{cn}$ (all symbols are as defined in the example in
Section~\ref{sec:NMR}):
\begin{eqnarray} 
	&   & \left[ \begin{array}{cccc}
	{\omega_a + \omega_b}	&{0}	&{0}	&{0} 	\\
	{0}	&{-\omega_b}		&{0}	&{0}	\\
	{0}	&{0}	&{- \omega_a + \omega_b}&{0}	\\
 	{0}	&{0}	&{0}	&{-\omega_b}		\end{array} \right]
\\
	& = & \omega_a ~\sigma_z \otimes |0 \rangle \langle 0| 
	+ \omega_b ~I \otimes \sigma_z
\label{eq:pureancilla}
\,.
\end{eqnarray}
The first term in Eq.(\ref{eq:pureancilla}) is the desired initial state.
The second term cannot affect the observable of interest, the spectrum at
$\omega_a/2\pi$, because of the following.
The identity in $A$ is invariant under the preparation pulse $Y_a(\theta)$.  
The input state is thus the identity, which has no coherence to start with.
Therefore, the output state after phase damping in both the control and the
coding experiment is still the identity.
This is non-trivial in the coding experiment.  However, inspection of 
Eqs.(\ref{eq:phi})-(\ref{eq:zzphi}) shows that spin $A$ is changed 
at most by a phase in the coding experiment.   
While Eqs.(\ref{eq:phi})-(\ref{eq:zzphi}) applies only to the case when $B$
starts in $|0 \rangle $, the result can be generalized to any arbitrary
diagonal density matrix in $B$ (proof omitted).
It follows from Eqs.(\ref{eq:ahigh})-(\ref{eq:blow})
that the second term is not observable in the output spectrum of $A$.

In contrast, the input state in spin $A$ can be a mixed state as given by the
first term in Eq.(\ref{eq:pureancilla}), since the phase damping code is still
applicable.  Different input states can be prepared by rotations about the
$\hat{y}$-axis of different angles $\theta \in [0,\pi]$ to span a semi-circle
in the Bloch sphere in the $\hat{x}\hat{z}$-plane.  Due to axisymmetry of
phase damping (Eq.(\ref{eq:phasedamp})), these states suffice to represent all
states to test the code.

We conclude with an alternative interpretation of the initial state
preparation.  Let the fractional populations of $|00 \rangle $, $|01 \rangle
$, $|10 \rangle $ and $|11 \rangle $ be $p_{00}$, $p_{01}$, $p_{10}$ and
$p_{11}$ in the thermal state.  Then, the initial state after temporal
labelling is
\begin{eqnarray} 
	&   & (p_{01} + p_{11})~I \otimes |1 \rangle \langle 1| 
	+ 2 \hspace*{0.5ex} p_{10}~I \otimes |0 \rangle \langle 0| 
\nonumber
\\	
	& & + ~2 \hspace*{0.5ex} (p_{00} - p_{10})~|00 \rangle \langle 00| 
\label{eq:initstate}
\,,   
\end{eqnarray}
where the identity term is not omitted, unlike Eq.(\ref{eq:pureancilla}). 
Temporal labelling serves to randomize spin $A$ in the first term in
Eq.(\ref{eq:initstate}) when $B$ is $|1 \rangle $.
We have shown previously that the identity input state of $A$ is preserved
throughout both the coding and the control experiments in the ideal case.
Consequently, only the last term in Eq.(\ref{eq:initstate}) contributes to any
detectable signals in all the experiments, and we can consider the last term
as the initial state.

Having justified both pictures using the first term in
Eq.(\ref{eq:pureancilla}) and the last term in Eq.(\ref{eq:initstate}) as
initial state, both will be used throughout rest of the paper.

\begin{figure}[ht]
\centerline{\mbox{\psfig{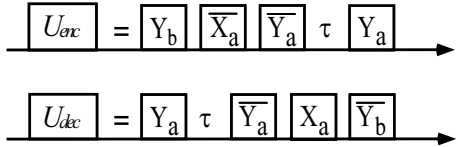}}}
\vspace*{1ex}
\caption{Pulse sequences to implement the encoder
	$U_{enc}$ and the decoder $U_{dec}$.  Time runs from left to right.
	$X(Y)_{a(b)}$ and $\tau$ are as defined in Section \ref{sec:NMR}.}
\label{fig:seqs}
\end{figure}

\hfill \\
\noindent {\bf Encoding and decoding}~~
The original encoding and decoding operations are composed of the Hadamard 
transformation and $C\!\!N_{ba}$, as defined in Section \ref{sec:NMR}.  
The actual sequences can be simplified and are shown in Fig.~\ref{fig:seqs}.

% FIG 7

The operator $U_{enc}$ can be found by multiplying the component
operators in Fig.~\ref{fig:seqs}, giving 
\begin{equation}
	U_{enc} = \frac{1}{\sqrt{2}} \left[ \begin{array}{cccc}

	{1}&{-1}&{0}&{0}
\\
	{0}&{0}&{i}&{i}
\\
	{0}&{0}&{1}&{-1}
\\
 	{i}&{i}&{0}&{0}

	\end{array} \right] 
\label{eq:uenc}
\,.
\end{equation}
The encoded states are slightly different from those 
in section~\ref{sec:2bitcode}:
\begin{eqnarray}
	|0_L \rangle & = &\frac{1}{\sqrt{2}} (|00 \rangle +i|11 \rangle )
\\
	|1_L \rangle & = &\frac{1}{\sqrt{2}} (i|01 \rangle +|10 \rangle )
\label{eq:codeexp}
\,, 
\end{eqnarray}
but the scheme is nonetheless equivalent to the original one. 

The decoding operation $U_{dec}$ is given by
\begin{equation}
	U_{dec} = \frac{1}{\sqrt{2}} \left[ \begin{array}{cccc}

	{1}&{0}&{0}&{-i}
\\
	{-1}&{0}&{0}&{-i}
\\
	{0}&{-i}&{1}&{0}
\\
 	{0}&{-i}&{-1}&{0}

	\end{array} \right] = U_{enc}^{\dagger}
\label{eq:udec}
\,. 
\end{equation}
The possible decoded outputs are the same as in Section~\ref{sec:2bitcode}
except for an overall sign in the single error cases.

\hfill \\
\noindent {\bf Storage}~~
The time delay between encoding and decoding corresponds to storage
time of the quantum state.  During this delay time, phase damping, amplitude
damping and $J$-coupling evolution occur simultaneously.  How to make phase
damping the dominant process during storage is explained as follows.

First of all, the time constants of amplitude damping, $T_1$'s, are much
longer than those of phase damping, $T_2$'s.  Storage times $t_d$ are chosen
to satisfy $t_d \leq T_2 \ll T_1$.  This ensures that the effects of amplitude 
damping are small.  

The remaining two processes, phase damping and $J$-coupling, can be considered
as independent and commuting processes in between any two pulses since all the
phase damping operators commute with the $J$-coupling evolution 
% 
% $\exp(-i \sigma_{z} \otimes \sigma_{z} \pi J t_d/2 )$.  
$\exp(-i \, \sigma_{z} \!\! \otimes \! \sigma_{z} \, \pi \! J t_d/2 )$.  
We choose $J t_d$ to be even integers to approximate the identity evolution.
As $J$ is known with limited accuracy, we add refocusing
$\pi$-pulses~\cite{Ernst94} to spin $B$ (about the $\hat y$-axis) in the
middle and at the end of the phase damping period to ensure trivial evolution
under $J$-coupling.  These pulses flip the $\hat{z}$ axis for $B$ during the
second half of the storage time so that evolution in the first half is always
reversed by that in the second half.  In this way, controllable amount of
phase damping is achieved to good approximation.

\hfill \\
\noindent {\bf Control Experiment}~~ 
For each storage time $t_d$, input state and temporal labelling experiment, a
control experiment is performed with the coding and decoding operations
omitted.  Since phase damping and $J$-coupling can be considered as
independent processes, and $J$-coupling is arranged to act trivially, the
resulting states illustrate phase damping of spin $A$ without coding.

\hfill \\
\noindent {\bf Output and readout}~~
For an input state prepared with $Y_a(\theta)$, the state after encoding,
dephasing and decoding ($\rho_5$ in Fig.~\ref{fig:scheme}) is derived in
Appendix~\ref{sec:mixstate} and is given by Eq.(\ref{eq:codedout})
(from here onwards, $\omega_a$ is omitted)
\begin{eqnarray}
	  \rho_5^{coded} = &\left[\rule{0pt}{2.1ex}\right. & 
		\cos\theta ~(1- p_a -p_b+2 p_a p_b)~ \sigma_z   
\nonumber
\\	  & 	& + \sin\theta ~(1- p_a -p_b)~ \sigma_x 
		\left.\rule{0pt}{2.1ex}\right] ~\otimes~ (I+\sigma_z)/2
\nonumber
\\
			+~& \left[\rule{0pt}{2.1ex}\right. & 
		\cos\theta ~( p_a +p_b-2 p_a p_b)~ \sigma_z  
\nonumber
\\	  & 	& + \sin\theta ~(- p_a  +p_b)~ \sigma_x 
		\left.\rule{0pt}{2.1ex}\right] ~\otimes~ (I-\sigma_z)/2
\label{eq:textcodedout}
\,. 
\end{eqnarray}
In the control experiment, the corresponding output state is given by 
(Eq.(\ref{eq:controlout}))
\begin{equation}
 	\rho_5^{control} = 
	\left[\rule{0pt}{2.1ex}\right. \cos\theta~\sigma_z + (1 - 2 p_a) 
	\sin\theta~\sigma_x \left.\rule{0pt}{2.1ex}\right] 
					\otimes (I+\sigma_z)/2
\label{eq:textcontrolout}
\,.
\end{equation}
The initial state used in the derivation of Eq.(\ref{eq:textcontrolout}) is
the first term in Eq.(\ref{eq:pureancilla}), and the encoding and decoding
operations are as given by Eqs.(\ref{eq:uenc}) and (\ref{eq:udec}).

In the ideal case, the output state can be read out in a single spectrum.
Recall that the coefficients of $-\sigma_y \otimes (I \pm \sigma_z)$ and
$-\sigma_x \otimes (I \pm \sigma_z)$ are the real and imaginary parts of the
low and the high frequency lines of $A$.  Therefore, the coefficients of
$-\sigma_z \otimes (I \pm \sigma_z)$ and $-\sigma_x \otimes (I \pm \sigma_z)$
in $\rho_5^{coded}$ and $\rho_5^{control}$ can be read out as the real and
imaginary parts of the low and the high frequency lines of $A$, if $X_a$ is
applied before acquisition.  This pulse transforms the $z$-component of $A$ to
the $y$-component leaving the $x$-component unchanged, as described in
Section~\ref{sec:NMR}.  Note that only states with spin $B$ being $|0 \rangle
$ ($|1 \rangle $) contribute to the low (high) frequency line.  Therefore, in
the coding experiments, the accepted (rejected) states of $A$ can be read out
separately in the low (high) frequency line.  There are no rejected states in
the control experiments.

The rest of the paper makes use of the following notation.  ``Output states''
or ``accepted states'' refer to the reduced density matrices of $A$ {\em
before} the readout pulse, and are denoted by $\rho_{a}^{coded} \equiv~_B
\langle 0|\rho_5^{coded}|0 \rangle _B$ and $\rho_{a}^{control} \equiv~_B
\langle 0|\rho_5^{control}|0 \rangle _B$.  Rejected states refer to $_B
\langle 1|\rho_5^{coded}|1 \rangle _B$ from the coding experiments.

The accepted and rejected states for a given input as calculated from
Eq.(\ref{eq:textcodedout}) and Eq.(\ref{eq:textcontrolout}) are summarized in
Table~\ref{table:outcome}.

% Table 1
\begin{table}[ht]
\begin{center}
{\small 
$	\begin{array}{|l|c|c|}
\hline
	{}&{z-\mbox{component} }&{x-\mbox{component} }
\\
\hline
\hline
	{\mbox{input state}} 
	& {\cos{\theta}} &{  \sin{\theta}} 
\\
\hline
	{\mbox{coding expt.:}} &{} &{}
\\
	{\mbox{accepted state}}
	&{\mbox{(1$-$p$_a$$-$p$_b$+2p$_a$p$_b$) cos$\theta$} } 
	&{\mbox{(1$-$p$_a$$-$p$_b$) sin$\theta$} } 
\\
	{\mbox{rejected state}}
	&{\mbox{ (p$_a$+p$_b$$-$2p$_a$ p$_b$) cos{$\theta$}} } 
	&{\mbox{ ($-$p$_a$+p$_b$) sin$\theta$}} 
\\
\hline
	{\mbox{control expt.:}} & {}&{}
\\
	{\mbox{accepted state}}
	&{\mbox{ cos$\theta$} } 
	&{\mbox{ (1-2p$_a$) sin$\theta$} }
\\	{\mbox{rejected state}}
	&{\mbox{0}}&{\mbox{0}} 
\\
\hline	
\end{array}
$
}
\end{center}
\vspace*{2ex} 
\caption{\label{table:outcome}Input and output states of spin $A$ 
in the coding and the control experiments.}
\end{table}

\begin{figure}[ht]
\centerline{\mbox{\psfig{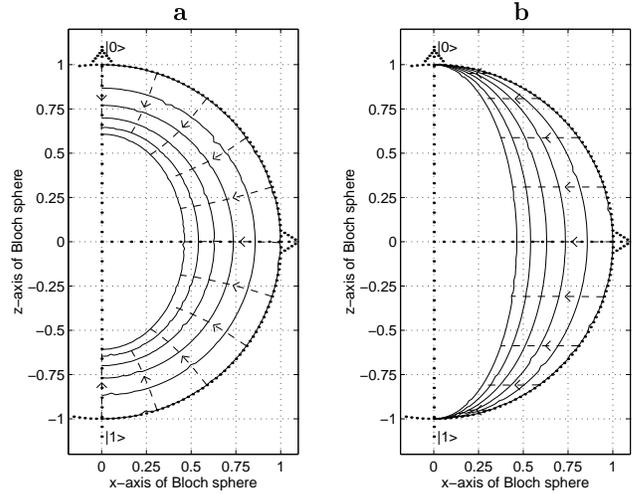}}}
\vspace*{1ex}
\caption{Predicted output states (a) with or (b) without coding.  The 
	arrows indicate the direction of time and the ellipses represent
	snapshots of the original surface of the Bloch sphere.}
\label{fig:prediction}
\end{figure}

The output states $\rho_{a}^{coded}$ and $\rho_{a}^{control}$, as predicted by
Table~\ref{table:outcome}, are plotted in Fig.~\ref{fig:prediction} in the
$\hat x \hat z$-plane of the Bloch sphere of $A$.  The north and south poles
represent the Bloch vectors $\pm \hat z$ ($|0 \rangle $ and $|1 \rangle $).
The time trajectories of various initial states are indicated by the arrows.
The Bloch sphere is distorted to an ellipsoid after each storage time.  We
concentrate on the cross-section in one half of the $\hat x \hat z$-plane, and
call the curve an ``ellipse'' for convenience.
The storage times plotted have equal spacing and correspond to $p_a = 0,
0.071, 0.133, 0.185, 0.230, 0.269$.  For each ellipse, $p_b$ is chosen to be
the same as $p_a$.  The main experimental results will comprise of information
of this type.

% FIG 8

\hfill\\
\noindent {\bf Fidelity}~~
One can quantify how well the input states are preserved using various
fidelity measures.  
In classical communication, the fidelity can be defined as the
probability of successful recovery of the input bit string in the worse case.
In quantum information processing, when the input is a {\em pure} state,
the above definition generalizes to the {\em minimal overlap fidelity},   
\begin{equation}
	{\cal F}  =  \mbox{min}_{\rho_{in}} \mbox{tr}(\rho_{out} \rho_{in})
\,.
\label{eq:fidelity1}
\end{equation}
We emphasize that Eq.(\ref{eq:fidelity1}) applies to {\em pure} input states
only.  For simplicity, fidelities for mixed input states will not be given
here.  The reason why Eq.(\ref{eq:fidelity1}) is sufficient for our purpose
will become clear later.
 
When $\rho_{in}$ and $\rho_{out}$ are qubit states of unit trace with
respective Bloch vectors $\hat{r}_{in}$ and $\vec{r}_{out}$, 
Eq.(\ref{eq:fidelity1}) can be rewritten as
\begin{equation}
	{\cal F} = \mbox{min}_{\hat{r}_{in}} \frac{1}{2} 
				(1 + \hat{r}_{in} \cdot \vec{r}_{out})
\label{eq:fidelity2}
\,.
\end{equation}
Recall from Eq.(\ref{eq:phasedampmix}) that, for phase damping, 
when $\hat{r}_{in} = (r_x, r_y, r_z)$, $\vec{r}_{out} =
(e^{-\lambda t} r_x, e^{-\lambda t} r_y, r_z)$.  Therefore, 
\begin{eqnarray}
	\hat{r}_{in} \cdot \vec{r}_{out} & = & e^{-\lambda t} 
	(r_x^2 + r_y^2) + r_z^2 
%\\
%	& = & - (1-e^{-\lambda t}) (r_x^2 + r_y^2) + 1
\\
	& = & - 2 p (r_x^2 + r_y^2) + 1 
\,,
\end{eqnarray}
where we have used the fact $|\hat{r}_{in}|^2 = 1$ for pure states and
$p=(1-e^{-\lambda t})/2$.  The minimum in Eq.(\ref{eq:fidelity2}) is
attained for input states on the equatorial plane with 
$r_x^2 + r_y^2 = 1$.  Therefore
\begin{equation}
	{\cal F} = 1 - p = \frac{1}{2} (1+e^{-\lambda t})
\label{eq:rate}
\,.
\end{equation}

With coding, the accepted state is (see Eqs.(\ref{eq:phi})-(\ref{eq:zzphi}))
\begin{equation}
	\rho_{a}^{coded} =  (1-p_a)(1-p_b) \rho_{in}
		 + p_a p_b \sigma_z \rho_{in} \sigma_z 
\,.
\end{equation}
If one considers the conditional fidelity in the accepted state, $\rho_{out}$
in Eq.(\ref{eq:fidelity1}) should be taken as the post measurement density
matrix, 
\begin{eqnarray}
	\rho_{out} & = & \frac{\rho_{a}^{coded}}{\mbox{tr}(\rho_{a}^{coded})} 
	= \frac{\rho_{a}^{coded}}{ (1-p_a)(1-p_b) + p_a p_b }
\\
	& \approx & (1-p_a p_b) \rho_{in} 
		   + p_a p_b \sigma_z \rho_{in} \sigma_z 
\,.  
\end{eqnarray}
Note that the above expression is identical to the expression for single qubit
phase damping but with error probability $p = p_a p_b$.  Therefore, coding
changes the conditional error probability to second order, and the conditional
fidelity is improved to ${\cal F}_C = 1-p_a p_b$.

The amount of distortion can also be summarized by the ellipticities of the
``ellipses'' that result from phase damping.  The ellipticity $\epsilon$ is
defined to be the ratio of the major axis to the minor axis.  Without coding,
the major axis remains unchanged under phase damping, and the minor axis
shrinks by a factor of $e^{-\lambda t}$, therefore $\epsilon = e^{\lambda t}$. 
Using Eq.(\ref{eq:rate}),
\begin{equation}
	{\cal F} = \frac{1}{2} (1+\frac{1}{\epsilon})
\,.
\label{eq:fande}
\end{equation}
With coding, ${\cal F}_C$ is given by the same expression on the right hand
side of Eq.(\ref{eq:fande}).  In the ideal case, the overlap fidelity and the
ellipticity have a one-to-one correspondence.  In the presence of
imperfections, the overlap fidelity and the ellipticity, one being the minimum
of the input-output overlap and the other being an average parameter of
distortion, are more effective in reflecting different types of distortion.

%%%%%%%%%%%%%%%%%%%%%%%%%%%%%%%%%%%%%%%%%%%%%%%%%%%%%%%%%%%%%%%%%%%%%%%%%%%%
%  new section for fidelities for deviation density matrices
%%%%%%%%%%%%%%%%%%%%%%%%%%%%%%%%%%%%%%%%%%%%%%%%%%%%%%%%%%%%%%%%%%%%%%%%%%%%

We now generalize to new definitions of fidelity for deviation density
matrices for the two bit code.
In NMR, quantum information is encoded in the small deviation of the state
from a completely random mixture.  
The problem with the usual definitions of fidelity is that they do not
change significantly even when the small deviation changes completely.
This is true whether the fidelities are defined for pure or mixed input
states.
% 
% The problem of the usual definition of fidelity as given in
% Eq.(\ref{eq:fidelity1}) is that it does not change significantly even 
% when the small deviation changes completely.
% 
To overcome this problem, we introduce the strategy of identifying the initial
excess population in $|00 \rangle $ as the pure initial state so that usual
definitions of fidelity for pure input states are applicable.  This improves
the sensitivity of the fidelity measures and provides a closer connection to
the pure state picture.

The initial state in Eq.(\ref{eq:initstate}) can be rewritten as
\begin{equation}
	\rho = \alpha \rho^{pure} + (1-\alpha) \rho^{quiet}
\,,
\end{equation}
where $\alpha = 2 (p_{00} - p_{10}) = \hbar \omega_a/2kT$, and  
\begin{eqnarray}
	\rho^{pure} & = & |00 \rangle \langle 00|
\\
	\rho^{quiet} & \approx & \frac{1}{1-\alpha} 
		\left[\rule{0pt}{2.1ex}\right. (p_{01} + p_{11})~\mbox{I} 
		\otimes |1 \rangle \langle 1| 
\nonumber \\
		& & \hspace*{10ex} + 2~p_{10}~\mbox{I} 
		\otimes |0 \rangle \langle 0| \left.\rule{0pt}{2.1ex}\right] 
\,.
\end{eqnarray}
It has already been shown that $\rho^{quiet}$ is irrelevant to the
evolution and the measurement of $\rho^{pure}$ when all processes are unital.
Therefore $\rho^{quiet}$ is neglected and the small signal resulting from the
slow non-unital processes will be treated as extra distortion to the 
observable component.
The input state prepared by $Y_a(\theta)$ can be written as  
\begin{equation}
	\rho_{in} = \alpha \rho_{in}^{pure} + (1-\alpha) \rho^{quiet}
\,.
\end{equation}
For the state change $\rho_{in} \rightarrow {\cal E}(\rho_{in})$, we consider
the overlap between $\rho_{in}^{pure}$ and ${\cal E}(\rho_{in}^{pure})$ in
place of the overlap between $\rho_{in}$ and ${\cal E}(\rho_{in})$.  
This defines a new overlap fidelity 
${\cal F}_\delta = \mbox{min}_{\rho_{in}^{pure}} 
		\mbox{tr}(\rho_{in}^{pure} {\cal E}(\rho_{in}^{pure}))$ 
$=$ $\mbox{min}_{\hat{r}_{in}} 
		\frac{1}{2} (1+\hat{r}_{in} \cdot \vec{r}_{out})$
similar to the pure state case.  

${\cal F}_\delta$ can be calculated from the experimental results in the
following manner.  The measured Bloch vector of $A$, $\vec{r}_m$, is
proportional to that defined by 
$_B \langle 0|{\cal E}(\rho_{in}^{pure})|0 \rangle _B$.  
Due to limitations in the measurement process, this proportionality constant 
$\tilde{\alpha}$ is not known a~priori.  
However, when $\theta=0$ in the control experiment, 
${\cal E}(\rho_{in}^{pure}) = \rho_{in}^{pure}$ and 
$\vec{r}_m = \tilde{\alpha} \hat{r}_{in}$.  Therefore, 
$\tilde{\alpha} = |\vec{r}_m|_{\theta = 0}$ can be determined.
In other words, $|\vec{r}_m|_{\theta=0}$ is used to normalize all other
measured output states before using the expression for ${\cal F}_\delta$.  

The expression for ${\cal F}_\delta$ can also be used for the conditional
fidelity in the coding experiment if the post-measurement accepted output
state is known.  This requires $\mbox{tr}(\rho_{a}^{coded})$ $=$ ($1-p_a-p_b
+2 p_a p_b$) to be determined for each storage time.  The correct
normalization is again given by the output at $\theta = 0$, which equals
$\vec{r}_m = \mbox{tr}(\rho_{a}^{coded}) \tilde{\alpha} \hat{r}_{in}$.

In summary, each ellipse obtained in the coding and the control experiment 
is normalized by the amplitude at $\theta=0$: 
\begin{equation}
	{\cal F}_\delta = \mbox{min}_{\hat{r}_{in}} \frac{1}{2} 
	\left[\rule{0pt}{2.4ex} 
	1+\frac{\hat{r}_{in} \cdot \vec{r}_{m}}{|\vec{r}_m(\theta=0)|}\right]
\,.
\label{eq:expfid}
\end{equation}
It is interesting to note that in contrast to the fidelity measure, the
ellipticity measure naturally performs an equivalent normalization, and thus
can be used for deviations without modifications.

We now turn to the experimental results, beginning with a description of our
apparatus.

%%%%%%%%%%%%%%%%%%%%%%%%%%%%%%%%%%%%%%%%%%%%%%%%%%%%%%%%%%%%%%%%%%%%%%%%%%%%%
% experiment and results 
%
\section{Apparatus and experimental parameters}
\label{sec:experiment}

\begin{figure}[ht]
\begin{center}
\mbox{\psfig{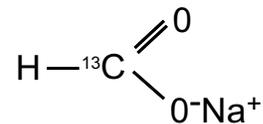}}
\end{center}
\vspace*{2.0ex}
\caption{
$^{13}$C-labeled formate.  The nuclear spins of the neighboring 
proton and carbon represent qubits $A$ and $B$. }
\label{fig:formate}
\end{figure}

We performed our experiments on carbon-13 labeled sodium formate
(CHOO$^-$Na$^+$) (Fig.~\ref{fig:formate}) at 15$^\circ$C.  The nuclear spins
of proton and carbon were used as input and ancilla respectively.  Note that
the system was heteronuclear.  The sodium formate sample was a 0.6 milliliter
1.26 molar solution (8:1 molar ratio with anhydrous calcium chloride) in
deuterated water \cite{chemicals}.  The sample was degassed and flame sealed
in a thin walled, 5mm NMR sample tube.

% FIG 9

The time constants of phase damping and amplitude damping are shown in
Table~\ref{table:t12}.  The fact $T_1 \ll T_2$ ensures that the effect of
amplitude damping is small compared to phase damping.  The experimental
conditions are chosen so that proton and carbon have almost equal $T_2$'s: the
most theoretically interesting scenario in which identical quantum systems are
available for coding.  Coding with qubits having very different $T_2$'s is
described in Appendix~\ref{sec:difft2}.

The time constants of phase damping and amplitude damping are shown in
Table~\ref{table:t12}.  
The fact $T_1 \ll T_2$ ensures that the effect of amplitude damping is small
compared to phase damping.
The experimental conditions are chosen so that proton and carbon have almost
equal $T_2$'s.  This eliminates potential bias caused by having a long-lived
ancilla when evaluating the effectiveness of coding.  This also realizes a 
common assumption in coding theory that identical quantum systems are
available for coding.  
Subsidiary experiments with qubits having very different $T_2$'s are 
described in Appendix~\ref{sec:difft2}.

% Table 2
\begin{table}[ht]
\centering
$\begin{array}{|c|c|c|}
\hline
	        &     T_1           & T_2               \\ \hline
   ~^1\mbox{H}~ & ~~9~\mbox{s}~~    & ~~0.65~\mbox{s}~~ \\ 
~^{13}\mbox{C}~ & ~~13.5~\mbox{s}~~ & ~~0.75~\mbox{s}~~ \\
\hline 
\end{array}
\vspace*{2ex}
$ 
\caption{\label{table:t12}
$T_1$'s and $T_2$'s for CaCl$_2$-doped formate at 15$^\circ$C, measured using
standard inversion recovery and Carr-Purcell-Meiboom-Gill pulse
sequences respectively.} 
\end{table}

Phase damping arises from constant or low-frequency non-uniformities of the
``static'' magnetic field which randomize the phase evolution of the
spins in the ensemble.
Several processes contribute to this inhomogeneity on microscopic or
macroscopic scales.
Which process dominates phase damping varies from system to 
system~\cite{Abragam}.  
For instance, intermolecular magnetic dipole-dipole interaction dominates
phase damping in a solution of small molecules, whereas the modulation of
direct electron-nuclear dipole-dipole interactions becomes more important if
paramagnetic impurities are present in the solution.  For molecules with
quadrupolar nuclei (spin $>$ 1/2), modulation of the quadrupolar coupling
dominates phase damping. 
Other mechanisms such as chemical shift anisotropy can also dominate phase 
damping in other circumstances. 
These microscopic field inhomogeneities have no net effects on the static
field when averaged over time, but they result in irreversible phase
randomization with parameters intrinsic to the sample.
Another origin of inhomogeneity comes from the macroscopic applied static
magnetic field.  In contrast to the intrinsic processes, phase randomization
due to this inhomogeneity can be reversed by applying refocusing pulses as
long as diffusion of molecules is insignificant.

Phase damping caused by the intrinsic irreversible processes alone has a time
constant denoted by $T_2$, while the combined process has a shorter time
constant denoted by $T_2^*$.  $T_2$ is measured by the
Carr-Purcell-Meiboom-Gill~\cite{Ernst94} experiment using multiple refocusing
pulses.  $T_2^*$ can be estimated from the line-width of the NMR spectral
lines: during acquisition, the signal decays exponentially due to phase
damping, resulting in Lorentzian spectral lines with line-width $1/ \pi
T_2^*$.

In our experiment, $T_2^*$'s for proton and carbon were estimated to within
0.05 s to be $\approx$ 0.35 and 0.50 s.  The storage times $t_d$ were
approximately 0, 62, 123, 185, 246, 308 ms ($n/J$ for $n=0,12,24,36,48,60$).
The maximum storage time was 120 $\tau$, long compared to the clock cycle and
was comparable to $T_2^*$.  The decay constant $\lambda$, defined in Section
\ref{sec:pd}, was given by $\lambda = 1/T_2^*$.  The error probabilities after
a storage time of $t_d$ were $p_i = (1-\exp(-t_d/T_{2i}^*))/2$ for spins
$i=A,B$.  To reconstruct the ellipse for each storage time, 11 experiments
were run with input states spanning a semi-circle in the $\hat x \hat
z$-plane.  Each input state was prepared by a $Y_a(\theta)$ pulse with $\theta
= n \pi/10$ for $n = 0,1,\cdots,10 $.

All experiments were performed on an Oxford Instruments superconducting magnet
of 11.7 Tesla, giving precession frequencies of $\omega_A/2\pi$ $\approx$ 500
MHz for proton and $\omega_B/2\pi$ $\approx$ 125 MHz for carbon.  A Varian
$^{\sf UNITY}${\sl Inova} spectrometer with a triple-resonance probe was used
to send the pulsed RF fields to the sample and to measure the FID's.  The RF
pulses selectively rotated a particular spin by oscillating on resonance with
it.  The $\pi$/2 pulse durations were calibrated, and they were typically 8 to
14 $\mu$s.  To perform logical operations in the respective rotating frames of
the spins, reference oscillators were used to keep track of the free
precession of both spins, leaving only the $J$-coupling term of 195.0 Hz in the
time evolution.  Each FID was recorded for $\approx$ 6.8 s (until the signal
had faded completely).  The thermal state was obtained after a relaxation time
of 80 s ($\gg T_1$'s) before each pulse sequence.

Using the above apparatus and procedures, we performed the experiments as
outlined in Section \ref{sec:2bitNMR}.  The experimental results are
described in the next section.  Tomography results following the 
evolution of the state are presented in Appendix~\ref{sec:tomo}.  

%%%%%%%%%%%%%%%%%%%%%%%%%%%%%%%%%%%%%%%%%%%%%%%%%%%%%%%%%%%%%%%%%%%%%%%%%%%
\section{Results and discussion}
\label{sec:result}

\subsection{Decoded Bloch spheres}
\label{sec:bloch}

The output states, $\rho_a^{coded}$ and $\rho_a^{control}$, obtained as
described in Sections~\ref{sec:2bitNMR} and \ref{sec:experiment}, and the
analysis that confirms the correction effects of coding, are presented in this
section.

\begin{figure}[ht]
\centerline{\mbox{\psfig{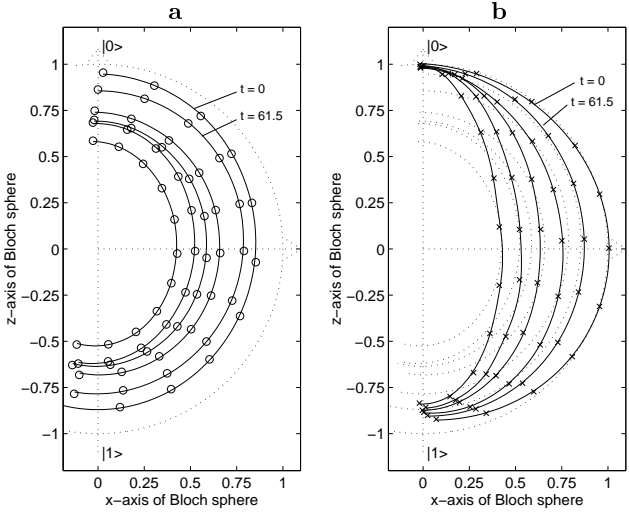}}}
\vspace*{2ex}
\caption{ 
(a) Experimental data (circles) showing the output states from the coding
experiment.  Each ellipse (solid line) corresponds to one storage time and is
obtained by a least-squares fit (Eq.(\ref{eq:intensity_mod})) to the data.
The storage times are $n \times 61.5$ms for $n=0,1,\cdots,5$, and smaller
ellipses correspond to increasing $n$.
(b) Experimental data (crosses) and fitted ellipses (solid lines) for the
control experiment.  A replica of figure (a) is plotted in dotted lines for
comparison.  In both figures, uncertainties in the data are much smaller 
than the circles and crosses.}
\label{fig:sodformellips}
\end{figure}

Fig.~\ref{fig:sodformellips} shows the accepted states in the
$\hat{x} \hat{z}$-plane of the Bloch sphere of $A$. $\rho_5^{coded}$ and
$\rho_5^{control}$ are plotted in Figs.~\ref{fig:sodformellips} a
and b.  
The ellipse for each storage time is obtained by a least-squares fit 
described later.  

% FIG 10

The most important feature in Fig.~\ref{fig:sodformellips} is the
reduction of the ellipticities of the ellipses due to coding, 
which represents partial removal of the distortion caused by phase
damping - the signature of error correction.  Coding is effective 
throughout the range of storage times tested.

We quantify the correction effects due to coding using the ellipticities.
When deviations from the ideal case such as offsets of the
angular positions of the points along the ellipses and attenuation of signal
strength with increasing $\theta$ exist, the minimum overlap fidelities and 
the ellipticities are no longer related by Eq.(\ref{eq:fande}).  
Since the ellipticity is an average measure of distortion which is less
susceptible to scattering of individual data points, we first study the 
ellipticities.  
Moreover, since the deviations from the ideal case are small, we can still
{\em infer} the fidelities from the ellipticities using Eq.(\ref{eq:fande}).
A discussion of the discrepancies and the exact overlap fidelities will given 
later.

\begin{figure}[ht]
\vspace*{0.5ex}
\centerline{\mbox{\psfig{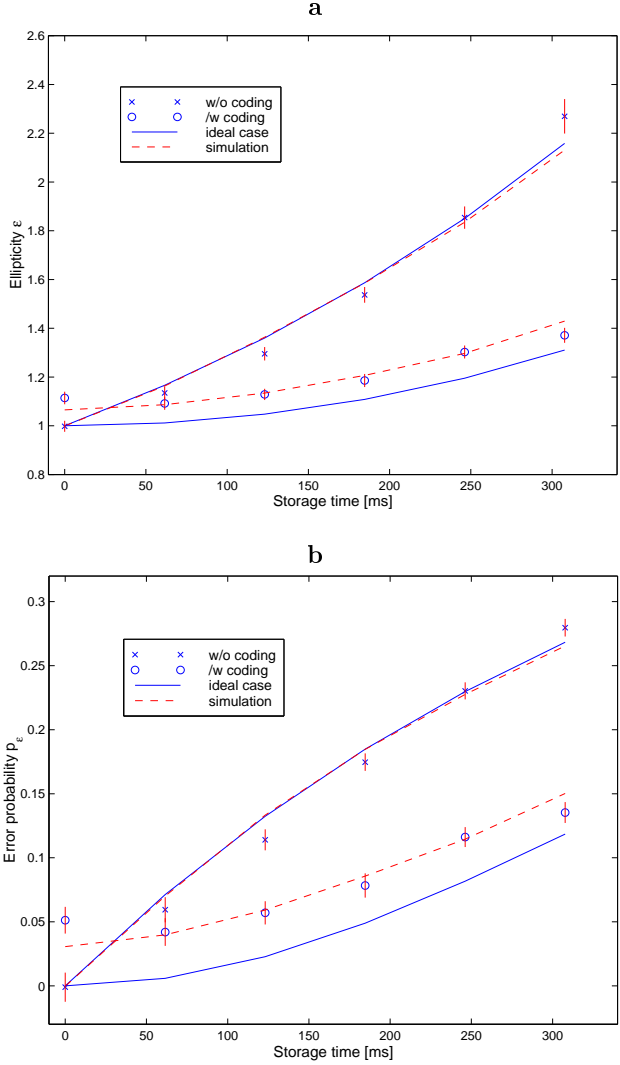}}}
\vspace*{2ex}
\caption{(a) Ellipticity and (b) inferred fidelity as a function of the 
	 storage time in the coding and the control experiments.  Error bars
	 represent 95\% confidence level.}
\label{fig:case1}
\end{figure}

\hfill\\
\noindent {\bf Ellipticities}~~
In the ideal case, the ellipticity for each ellipse can be obtained
experimentally as
\begin{equation}
	\epsilon =  \sqrt{\frac{I(\theta = 0)}{I(\theta=\frac{\pi}{2})}}
\label{eq:epsilon}
\,,
\end{equation}
where $I$ denotes the intensity (amplitude square) of the peak integral.  $I$
is given by the $\hat x$ and $\hat z$-components of the output states as
\begin{equation}
	I = r_x^2 + r_z^2
\,. 
\end{equation}
In the ideal case, $I(\theta)$ can be found from Table~\ref{table:outcome}:
\begin{eqnarray}
	I_{control}(\theta) & = & 1 - 4 (p_a - p_a^2) \sin^2\theta
\\
	I_{coded}(\theta) & = & (1 - p_a - p_b + 2 p_a p_b )^2
\nonumber
\\
	& & - 4 p_a p_b (1- p_a - p_b + p_a p_b) \sin^2\theta
\,, 
\end{eqnarray}
and both are of the functional form 
\begin{eqnarray}
	I_{ideal}(\theta) & = & A + B \sin^2\theta  
\label{eq:intensity}
\,. 
\end{eqnarray}
Experimentally, the output Bloch vectors do not form perfect ellipses.  We
modify Eq.(\ref{eq:intensity}) to include signal strength attenuation with
increasing $\theta$ and constant offsets in the angular positions:
\begin{eqnarray}
      I_{exp}(\theta) & = & (A + B \sin^2(\theta + D)) (1- C (\theta + D))
\label{eq:intensity_mod}
\,, 
\end{eqnarray}
and perform non-linear least-squares fits of the experimental data to
determine $A,B,C$ and $D$.  The fitted ellipses plotted in
Fig.~\ref{fig:sodformellips} follow from Eq.(\ref{eq:intensity_mod}) and the
fitted parameters.
The ellipticities $\epsilon$ are found using Eq.(\ref{eq:epsilon}) by
interpolating the intensities at $\theta = 0$ and $\theta = \frac{\pi}{2}$.
The ellipticities are plotted in Fig.~\ref{fig:case1}~a.
The uncertainties of the fitted parameters originate from the uncertainties in
the data, which are estimated to be $\approx$ 1\% for the amplitude and 1.5
degrees for the phase in the measured peak integrals.  These uncertainties are
propagated numerically to the ellipticities as plotted in
Fig.~\ref{fig:case1}~a.
Ideal case predictions and simulation results are also plotted in
Fig.~\ref{fig:case1}~a.  The simulation takes into account the major
imperfection in the pulses and will be described later.
%

% FIG 11

\hfill \\
\noindent {\bf Error correction}~~
The effectiveness of coding to correct errors is evident when comparing 
the ellipticities from the coding and the control experiments 
(Fig.~\ref{fig:case1}~a).
Without coding, the ellipticity grows exponentially as $e^{t_d/T_{2a}^*}$
(for $T_{2a}^*$ fitted to be $\approx 0.4s$).  
With coding, the growth is slowed down, with almost zero growth for small 
$t_d$.
The suppression of linear growth of the ellipticity can be further quantified
by weighted quadratic fits $\epsilon = c_0 + c_1 t_d + c_2 t_d^2$ to the 
ellipticities.  
% 
% For the control experiment, $c_0 = 1.00 \pm 0.01$, $c_1 = 1.31 \pm 0.21$ 
% and $c_3 = 8.81 \pm 0.82$ whereas for the 
% the coding experiment: $c_0 = 1.10 \pm 0.02$, $c_1 = -0.24 \pm 0.29$ 
% and $c_3 = 3.80 \pm 0.91$ ($t_d$ is in seconds).  
%
For the control experiments, $c_0 = 1.00 \pm 0.01$, $c_1 = 1.31 \pm 0.21$ 
and $c_3 = 8.8 \pm 0.8$ whereas for the 
the coding experiments, $c_0 = 1.10 \pm 0.02$, $c_1 = -0.24 \pm 0.29$ 
and $c_3 = 3.8 \pm 0.9$ ($t_d$ in seconds).     
Therefore, the linear term ``vanishes'' due to coding.  The small negative
coefficient for the linear term originates from the scattering of the data
point at zero storage time.

To quantify the ``cost of the noisy gates'' caused by the imperfect pulses, we
compare the ellipticities from the coding experiments and from the ideal case,
the quadratic fits of which are respectively $\epsilon^{coded}_{expt} = 1.10 -
0.24 t_d + 3.80 t_d^2$ and $\epsilon^{coded}_{ideal} = 1.00 + 0.15 t_d + 2.50
t_d^2$.  The imperfections cause the ellipticity to increase by 0.1 at $t_d =
0$ and this extra distortion {\em decreases} with $t_d$.
We take advantage of the fact that the simulation results are close to the
data points but are not as scattered to have a better estimate of this ``cost
of the noisy gates''.  The simulation data can be fitted by
$\epsilon^{coded}_{sim} = 1.06 + 0.32 t_d + 2.47 t_d^2$.  Compared to the
ideal case, the coding operations increase the ellipticity by $\approx$ 0.06
at $t_d = 0$, and this extra distortion remains almost constant for all $t_d$.

The error probabilities as inferred from the ellipticities $p_\epsilon = 1 -
{\cal F}_\epsilon = \frac{1}{2}(1-\frac{1}{\epsilon})$ are plotted in
Fig.~\ref{fig:case1}~b as a function of storage time.

\begin{figure}[ht]
\centerline{\mbox{\psfig{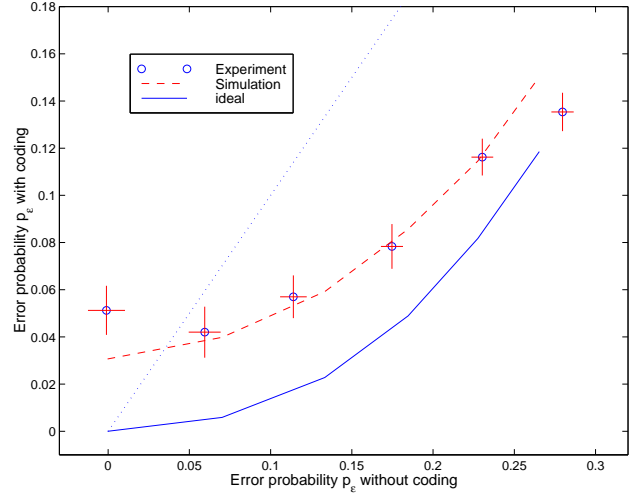}}}
\vspace*{2ex}
\caption{Error probabilities in the coding experiments vs the corresponding
	 values in the control experiments.  Error bars represent 95\%
	 confidence level.  The 45$^{\circ}$ line is plotted as a dotted
	 line. }
\label{fig:eorder}
\end{figure}

Error correction is also manifested by expressing $p_\epsilon$ in the coding
experiment as a function of the original $p_\epsilon$ in the control
experiment, as plotted in Fig.~\ref{fig:eorder}.
The quadratic fit to the experimental results gives 
$p^{coded}_{exp} = c_0 + c_1 p + c_2 p^2$ where 
$p$ stands for $p_\epsilon$ in the control experiment, 
$c_0 = 0.047 \pm 0.008$, $c_1 = -0.05 \pm 0.12$ and $c_2 = 1.38 \pm 0.40$.  
Therefore, the expected improvement $p \rightarrow p_a p_b$ is confirmed.
%
% The curve fits for the ideal case gives $p^{coded} = - 0.032 p + 1.783 p^2$.
%
Experimentally, the error probabilities are larger than in the ideal case 
by at most 4.7\% and these extra errors decrease with $p$.
The quadratic fit to the simulation results (which is a good approximation of
the experimental data) gives $p^{coded}_{sim} = 0.032 - 0.032 p + 1.783 p^2$ 
and differs from the ideal case by a constant amount of $\approx 0.033 \pm
0.003$ for all $p$, which represents the cost of the noisy gates.  In
conclusion, coding with noisy gates is still effective in our experiments,
even though the noisy gates add a constant amount of distortion.

% FIG 12

%%%%%%%%%%%%%%%%%%%%%%%%%%%%%%%%%%%%%%%%%%%%%%%%%%%%%%%%%%%%%%%%%%%%%%%%%
\subsection{Discrepancies}
\label{sec:discrep}

While the data exhibit a clear correction effect, there are notable deviations
from the ideal case.
First, the ellipses with coding are smaller than their counterparts without 
coding.
This is most obvious when the storage time is zero, in which case the coding
and the control experiments should produce equal outputs.
Second, the signal strength is attenuated with increasing $\theta$ relative to
ideal ellipses.  Third, although the data points are well fitted by ellipses,
their angular positions are not exactly as expected (``$\theta$-offsets'').
Finally, the spacings between the ellipses deviate from expectation.  The
causes of these discrepancies and their implications on error correction are
discussed next.

\hfill \\
\noindent {\bf Gate imperfections: RF field inhomogeneity}~~
%We find from other experiments that 
The major cause of experimental errors is RF field inhomogeneity, which causes
gate imperfections.  This was determined by a series of experiments (details
of which are not given here), and a thorough numerical simulation, as
described below.  The physical origin of the problem is as follows.
The coil windings produce inhomogeneous RF fields that randomize the angles of
rotation among molecules.  For a single rotation, the signal averaged over the
ensemble decreases exponentially with the pulse duration to good
approximation.  A measure of the RF field inhomogeneity is the signal strength
after a $\pi/2$ pulse.  They are measured to be $\approx$ 0.96 and 0.92 for
proton and carbon respectively.
%(0.956 and 0.920 for proton and carbon in the Formate experiment)
%(0.958 and 0.920 in the Chloroform experiment).
In other words, a single $\pi/2$ pulse has an error of $\approx$ 4-8\%.

RF field inhomogeneity affects our experiments in many ways.  
First, it attenuates the signal in both the coding and the control
experiments, but the effects are much more severe in the coding experiments
which have eight extra pulses.  For instance, when the storage time is zero,
the two experiments should have identical outputs, but the ellipse in the
coding experiment is actually 5-15~\% smaller.  Second, for each ellipse,
attenuation increases with $\theta$ as the preparation pulse $Y_a(\theta)$
becomes longer.

The effects of the RF field inhomogeneity are complicated, because the errors
from different RF pulses are correlated, and the correlation depends on the
temporal separation between the pulses and the diffusion rate of the
molecules.  The correlation time of the RF field inhomogeneity is comparable
to the experimental time scales.  For this reason, predictions of the effects
of RF field inhomogeneity are analytically intractable.

\begin{figure}[ht]
\centerline{\mbox{\psfig{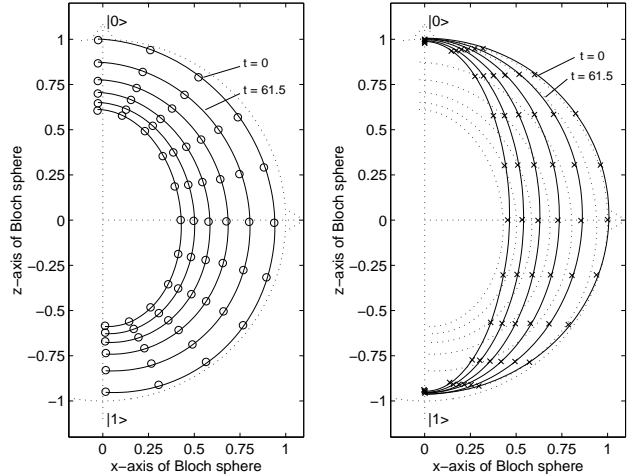}}}
\vspace*{2ex}
\caption{Simulated output states, plotted similarly as in 
	Fig.~\ref{fig:sodformellips}.  The simulation results are fitted 
	by ellipses similar to those of the experimental data.}
\label{fig:simellips}
\end{figure}

Numerical simulations were performed to model the dominant effects of RF field
inhomogeneity.  We followed the evolution of the states assuming random RF
field strengths drawn from Lorentzian distributions (also known as Cauchy
distributions) with means and standard deviations matching pulse calibration
and attenuation for the $\pi/2$ pulses.  All parameters in the simulation,
including $T_2^*$'s, were determined experimentally without introducing any
free parameters.  As the exact time correlation function for the errors was
unknown, except for numerical evidence of a long correlation time, we {\em
assumed} perfect correlation in the errors.  The simulated ensemble output
signal was obtained by Gaussian integration with numerical errors bounded to
below 1.5\%.  The results were shown in Fig.~\ref{fig:simellips}.

% FIG 13

Besides phase damping and error correction effects in the data, the
simulations also reproduce extra signal attenuation in the coding experiments.
The ellipticities obtained from simulations (see Fig.~\ref{fig:case1})
approximate the experimental values very well.  Such agreement to experimental
results is surprising in the absence of free parameters in the model.
Simulation results allow the discrepancy between the observed and the ideal
ellipticities to be explained in terms of the RF field inhomogeneity and allow
the ``cost of coding'' to be better estimated to be the constant 6~\% increase
in ellipticity or the $\approx$ 3~\% increase in error probabilities.

The simulation results also predict increasing attenuation with $\theta$.
From the fitted parameter $C$ (see Eq.(\ref{eq:intensity_mod})), the
amplitudes at $\theta = \pi$ are $\approx$ 4\% {\em weaker} than the
corresponding values at $\theta =0$ in the simulations.  Experimentally, this
attenuation increases from $\approx$ 8 to 15\% (as storage time increases from
0 to 308 ms) in the control experiments, and remains $\approx$ 8\% in the
coding experiments.  Therefore, RF field inhomogeneity contributes to the
attenuation but only partially.

We conclude that RF field inhomogeneity as we have modeled explains the
diminished signal strength in the coding experiments.  The simulation
quantifies the ``cost of the noisy gates''.  RF field inhomogeneity also
explains part of the attenuation with increasing $\theta$.  We can also
conclude that other discrepancies not predicted by the simulations are {\em
not} caused by RF field inhomogeneity and these discrepancies are
described next.

\hfill \\
\noindent{\bf Other discrepancies}~~
The simulation results show that RF field inhomogeneity does {\em not} explain
why the attenuation at large $\theta$ increases with storage time without
coding, and it does not explain the $\theta$-offsets along the ellipses and
the unexpected spacings between them.

The increased attenuation with storage time at large $\theta$ can be caused by
amplitude damping.  A precise description \cite{ad} of amplitude damping
during storage is out of scope of this paper and we consider only the dominant
effect predicted by a simple picture.  {\em Phenomenologically}, the loss of
energy to the lattice is described in the NMR literature by
\begin{equation}
	z(t) = z(\infty) + (z(0) - z(\infty))~e^{-t/T_1}
\,,   
\end{equation}
where $z(\infty)=1$ is the thermal equilibrium magnetization.  As $z(0) = \pm
1$ at $\theta = 0$ and $\pi$, we expect no changes at $\theta = 0$, 
but expect $|z|$ at $\theta = \pi$ to decrease by $0-7$~\% for 
$t_d \approx 0-308$ ms and $T_1 \approx 9$ s for proton.
Note that refocusing does not affect spin $A$ in the control
experiments~\cite{refocusing} but it swaps $|0_L \rangle $ and $|1_L \rangle $
halfway during storage, symmetrizing the amplitude damping effects in the
coding experiments.  Therefore, we expect increased attenuation with storage
time in the control experiments only.  This matches our observations that
$I(\theta = \pi)$ decreases from 8 to 16~\% in the control experiments, and
remains 8\% in the coding experiments.  Moreover, earlier data taken without
refocusing (not presented) have the same trend of increased attenuation with
storage time in both the coding and the control experiments.  These are all in
accord with the hypothesis that amplitude damping is causing the observed
effect.

The second unexplained discrepancy is that the output states span more than a
semi-ellipse in the coding experiment but slightly less than a semi-ellipse in
the control experiment.  We are not aware of any quantum processes that can be
a cause of it.  
It is notable that the output states and the fitted ellipticities can be used
to infer the initial values of $\theta$, and they are roughly proportional to
the expected values for each ellipse.  The proportionality constants are 5-8
\% higher than unity in the coding experiment, and 0-1.6\% lower in the
control experiments.  Moreover, similar effects are observed in many other
experimental runs.
Therefore, this is likely to be a systematic error.

We have no convincing explanation for the anomalous spacings between the
ellipses in the experiment.
However, from the fact that all the data points belonging to the same ellipse
are well fitted by it, the anomalous spacings are unlikely to be caused by
random fluctuations on the time scales of each ellipse-experiment.
The effect of the anomalous spacings is reflected in the scattering of the
ellipticities of the data, and the large uncertainties in the quadratic fits.

While it is impossible to eliminate or to fully explain these imperfections, 
it is possible to show that the deviations cannot affect the conclusion that
error correction is effective.

\hfill \\
\noindent {\bf Effects of the discrepancies}~~
We now consider the effects of the discrepancies on the ellipticities and the
inferred fidelities in the experiments.  First of all, radial attenuation of
the signal due to RF field inhomogeneity does not affect the ellipticities nor
the inferred fidelities (taken as conditional fidelities).  
Second, different expressions for the ``ellipticity'' are not equivalent when
the output states do not form perfect ellipses.  However, they differ by no
more than 7 and 3~\% in the control and the coding experiments.
$\theta$-offsets along the ellipses are not reflected in the ellipticities,
resulting in overestimated inferred fidelities.  This is bounded by
3\%.   
The scattering of the ellipticities due to anomalous spacings between the 
ellipses is averaged out with curve-fits to the data.  
The most crucial point is, none of these effects have a dependence on the
storage time that can be mistaken as error correction.  Therefore, the effects
of error correction can still be confirmed in the presence of all these small
discrepancies.

\subsection{Overlap fidelity}

The two previous subsections dealing with the ellipticities provide an
analysis of the global performance of the code.  A stricter analysis is
provided in this section using the overlap fidelities given by
Eq.(\ref{eq:expfid}) in Section~\ref{sec:2bitNMR}. 
The minimal overlap reflects all defects and deviations hidden in the
ellipticities as well as other distortions such as that caused by amplitude
damping.

\begin{figure}[ht]
\centerline{\mbox{\psfig{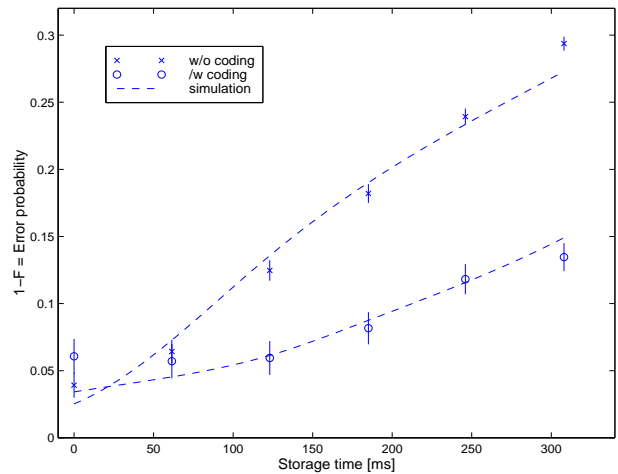}}}
\vspace*{1ex}
\caption{Overlap fidelity as defined in Eq.(\ref{eq:expfid}).  
Points indicate experimental data and dashed lines indicate 
simulation results.  Error bars represent 95\% confidence level.}
\label{fig:rawfidelity}
\end{figure}

All measurements are normalized using the amplitudes at $\theta=0$ as
described in Section~\ref{sec:2bitNMR}.  In the control experiments, the
output states at $\theta=0$ are least affected by amplitude damping and RF
field inhomogeneity.  Therefore, the normalization can be done accurately. 
In the coding experiment, the signal attenuation at $\theta = 0$ due to RF
field inhomogeneity can lead to overestimated fidelities.  We determine the
uncertainties due to RF field inhomogeneity by the following method.
For each storage time, the amplitudes of the accepted and the rejected states
at $\theta = 0$ are summed.  The sum is compared with the corresponding
amplitude at $\theta = 0$ in the control experiment to estimate the
attenuation due to RF field inhomogeneity.  The effects on the overlap
fidelities are bounded to below 2\%.  
The errors in the measured peak integrals are propagated to the fidelities 
which result in standard deviations no more than 0.7\%.
We apply similar procedures to the simulation results.   
The net error probabilities, given by $1-{\cal F}$ for the control and
$1-{\cal F}_C$ for the coding experiments, are plotted in
Fig.~\ref{fig:rawfidelity}.  

% FIG 14

The large difference in the rates of growth of error probabilities confirms 
the effectiveness of coding even when a stricter measure of fidelity is used.  

%%%%%%%%%%%%%%%%%%%%%%%%%%%%%%%%%%%%%%%%%%%%%%%%%%%%%%%%%%%%%%%%%%%%%%%%%%%
\section{Conclusion}
\label{sec:conclusion}

We have demonstrated experimentally, in a bulk NMR system, that using a two
bit phase damping detection code, the distortion of the accepted output states
can be largely removed.
These experimental results also provided quantitative measures
of the major imperfections in the system.  The principle source of errors,
RF field inhomogeneity, was studied and a numerical simulation was developed
to model our data.  Despite the imperfections, a net amount of error
correction was observed, when comparing cases with and without coding, and
including gate errors in both cases.

Our analysis also addresses several theoretical questions in quantum error
correction in bulk samples such as the fidelity measures of deviation density
matrices.  In the following, we conclude with some observations regarding
quantum error correction in bulk systems, including syndrome measurements, the
equivalence between error correction and logical
labelling~\cite{Gershenfeld97,Chuang97e}, the applicability and advantages of
detection codes and some issues in signal strength.

Projective syndrome measurements traditionally employed in the standard
theory of quantum error correction are impossible in ensemble quantum
computation.  Measurements via the acquisition of the FID do not reduce
individual quantum states and provide only ``average syndromes''.  Moreover,
the quantum states are destroyed after acquisition.  However, the important
point is that syndrome measurement is {\em not necessary} in error
correction~\cite{Peres96,Aharonov96}.

In each molecule, the syndrome bits carry the error syndrome for that
particular molecule after decoding.  These bits can either be used in a
controlled-operation to correct the error~\cite{Peres96}, or in the case of a
detection code, to ``logically label'' the correct and erroneous states.
Conversely, logical labelling to obtain effective pure states can be
considered as error detection: unsuitable initial states are ``detected'' and
are labelled as ``bad'' to start with.  Both processes involve ejecting the
entropy of the system to the ancilla bits.  Error detection and logical
labelling are therefore similar concepts.

The distinction between correction and detection codes is blurred when using
bulk systems.  The objective of error correction is to achieve reliable data
transmission or information processing with high probability of success.  When
information is encoded in single systems, encoding with an error detection
scheme will fail to provide reliable output for two different reasons:
accepting an erroneous state or losing the state upon the detection of an
error.  Therefore, coding schemes capable of {\em distinguishing and
correcting} errors are necessary to improve the probability of successful data
processing.  In contrast, in a bulk sample, a large initial redundancy exists
upon preparation, and the combined signal of all the accepted cases forms the
output.  Therefore, rejection of the erroneous cases results in a reduction
of the signal strength in the improved accepted cases without necessarily
causing a failure.  Detection codes thus provide a tradeoff between
probability of error-free computation and signal strength.

As suggested by this analysis, and in concert with our experimental results,
it thus makes sense to use detection codes instead of correction codes in
bulk quantum computing systems under certain circumstances.  Fundamentally,
it is valuable to be able to interchange resources depending on their
relative costs, This is illustrated by the following simple example.
Suppose a total pool of $m$ qubits is available for transmission, and
one just wants to correct for single phase flip errors of probability
$p$.  Using a three bit code, one would obtain an aggregate signal
strength of $m/3$, with fidelity $1-3p^2$, whereas with a
two bit code, the accepted signal strength would be $m(1-2p)/2$, with 
fidelity $1-p^2$.  Therefore, when $p\leq 1/6$, the two-qubit
code performs better in this model due to its higher rate. 

Another example relevant to bulk computation arises when the encoding and
decoding circuits fail with probability proportional to the number of
elementary gates used.  Although errors in consecutive gates can be made to
cancel sometimes, this basic scenario is substantiated by our experiment, in
which imperfect pulses contribute significantly to the net error.  Assume now
that we have $n$ molecules, which are either two or three qubit systems. 
Let us compare the performance of the two and three bit codes, based on the
strength of the correct output signal.  Because the correction code requires
at least three times as many operations as the detection code~\cite{gates},
the figures of merit obtained for the two schemes are $n (1-3 p_g)$ and $n
(1-2p)(1-p_g)$ respectively, where $p_g$ is the gate failure probability.  In
this model, the detection code performs better for $p \leq p_g/(1-p_g)$ 
due to the simplicity of the coding operations.

A third example is the case of current state NMR quantum computation at room
temperature, in which the intrinsic signal strength decreases exponentially
with the number of qubits\cite{Gershenfeld97,Chuang97e,Knill97}.  In this
model, the initial signal strength of an effective pure state of $m$ qubits is
approximately of order $2^{-m}$, and thus, for an ensemble of $n$ molecules,
the signal strengths of the outputs from the correction and detection codes
are about $n/8$ and $n(1-2p)/4$ respectively.  According to this measure of
performance, the detection code outperforms the correction code for $p \leq
1/4$ ($p\leq 0.27$ in our experiments).

If signal strength indeed decreases exponentially with $m$, then some
interesting generalizations can be made.  For arbitrary qubit errors, a
$t$-error detection code has distance $d \geq t+1$, while a $t$-error
correction code has distance $d \geq 2t+1$~\cite{Gottesman97}.  If one
encodes $k$ bits in $l$, the extra number of qubits used, $l-k$, satisfies
the singleton bound~\cite{Knill96,Cerf,Calderbank}, $l-k \geq 2d-2$.
Therefore, the output signal strengths for detection and correction codes
would be approximately proportional to $(1-p f(p))/2^{2t}$ and $1/2^{4t}$,
where $f(p)$ is a polynomial.  The detection code is thus always better
asymptotically in this model\cite{despite}.

This work illustrates how a careful study of dynamics in bulk quantum systems
can provide a valuable opportunity to demonstrate and test theories of quantum
information and computation.  The development of temporal, spatial, logical,
and related labeling techniques opens a window allowing information about the
dynamics of single quantum systems to be extracted from bulk systems.
Furthermore, by systematically developing an experimental toolbox of quantum
circuits and quantum error correction and detection codes, experiments which
test multiple particle quantum behavior become increasingly accessible.  With
improvements in the initial polarization in the systems and new labeling
algorithms which do not incur exponential signal loss~\cite{Schulman98}, and
with better methods to control the major source of error, the RF field
inhomogeneity, we believe that further study of bulk quantum systems will
complement the study of single quantum systems, provide new insights into the
dynamical behavior of open quantum systems, and further the potential for
quantum information processing.

%%%%%%%%%%%%%%%%%%%%%%%%%%%%%%%%%%%%%%%%%%%%%%%%%%%%%%%%%%%%%%%%%%%%%%%%%%%
\section{Acknowledgments}

This work was supported by the DARPA Ultrascale Program under contract
DAAG55-97-1-0341.  D.L. acknowledges support of an IBM Graduate Fellowship.
L.V.  acknowledges a Yansouni Family Fellowship under the Stanford Graduate
Fellowship program.  L.V., X.Z. and D.L. are indebted to Prof. James Harris
and Prof. Yoshihisa Yamamoto for their patience and support.  We thank
Thorsten Hesjedal, Lois Durham and Jody Puglisi for experimental assistance,
Nabil Amer, Charles Bennett, Hoi Fung Chau, Hoi-Kwong Lo and Shigeki
Takeuchi for enjoyable and helpful discussions, and Rolf Landauer for his
guiding wisdom.

%%%%%%%%%%%%%%%%%%%%%%%%%%%%%%%%%%%%%%%%%%%%%%%%%%%%%%%%%%%%%%%%%%%%%%%%%%%%%
% References
 
% \bibliographystyle{prsty}
% \bibliography{qc}

\begin{thebibliography}{10}
\bibitem{chemicals} Formic-13C acid and calcium chloride were 
supplied by Aldrich (respective catolog Nos. 27,941-2 and 23,922-4). 
Chemicals were used without further purification.
\bibitem{ad} We note that $J$-coupling, phase damping and amplitude damping
are non-commuting and simultaneous processes.  Moreover, the effects of
generalized amplitude damping at finite temperature on deviation density
matrices are complicated.
\bibitem{refocusing} The refocusing pulses were applied to spin $B$, 
therefore spin $A$ was not affected directly.  Refocusing affected the output 
indirectly by flipping spin $B$ for half of the storage time during 
which it was susceptible to amplitude damping.
\bibitem{gates} Two hadamard transformations and two controlled-nots are
needed in the detection code, whereas six hadamard transformations, four
controlled-nots and one Toffoli gate are needed in the correction code.
Simulating the Toffoli gate by three controlled-nots, the correction code
still requires at least three times as many operations as required by the
detection code.
\bibitem{chloroform} The chloroform sample was a 0.5 milliliter, 0.2 molar
solution in d6-acetone supplied by Cambridge Isotope Labs, Inc. 
(catolog No. CLM-262).  Experiments were performed at room temperature.
\bibitem{chcl3t2} $T_2^*$'s for the input and the ancilla, as measured from
the line widths, were 0.13 and 0.53 s when carbon was the input, and 0.92s and
0.16 s when proton was the input.  The storage times were 0, 37.2, 74.4,
111.6, 148.8 ms when carbon was the input, and 0, 223, 446 ms when proton
was the input.
\end{thebibliography}

%%%%%%%%%%%%%%%%%%%%%%%%%%%%%%%%%%%%%%%%%%%%%%%%%%%%%%%%%%%%%%%%%%%%%%%%%%%
\appendix

\begin{onecolumn}

\section{Mixed state description of the two bit code}
\label{sec:mixstate}

Recall that the initial state after ancilla preparation is given by
$\rho_0=\sigma_z \otimes (I+\sigma_z)/2$ (see Eq.(\ref{eq:pureancilla}), with 
$\omega_a$ omitted).
After $Y_a(\theta)$, the new density matrix is given by 
\begin{equation}
   \rho_1 = (\cos\theta \sigma_z + \sin\theta \sigma_x) \otimes (I+\sigma_z)/2
\,.
\end{equation}

Without coding, phase damping changes the density matrix to 
\begin{equation}
	\rho_5^{control} = \left[\rule{0pt}{2.4ex} 
	\cos\theta \sigma_z + (1 - 2 p_a) \sin\theta 
	\sigma_x \right] \otimes (I+\sigma_z)/2	  
\label{eq:controlout}
\,.
\end{equation}

With coding, the encoding, phase damping, and decoding
change the density matrix to $\rho_3$, $\rho_4$, and $\rho_5$: 
\begin{eqnarray}
	\rho_3^{coded} 	& = & 
	\cos\theta (\sigma_z \otimes \sigma_z + \sigma_y \otimes \sigma_x)/2  
\nonumber
\\
	& + & \sin\theta (\sigma_x \otimes I + I \otimes \sigma_y)/2  
\\
	\rho_4^{coded} 	& = &  
	\cos\theta (\sigma_z \otimes \sigma_z + 
		    (1-2 p_a) (1-2 p_b) \sigma_y \otimes \sigma_x)/2    
\nonumber
\\
	& + & \sin\theta ((1-2 p_a) \sigma_x \otimes I + 
			  (1-2 p_b) I \otimes \sigma_y)/2  
\\	
	\rho_5^{coded} 	& = &  
	\cos\theta \sigma_z \otimes (I + (1-2 p_a) (1-2 p_b) \sigma_z)/2
\nonumber
\\
	& + & \sin\theta \sigma_x \otimes ((1-2 p_a) I + (1-2 p_b) \sigma_z)/2 
\\
	& = &  \cos\theta \sigma_z \otimes \left[\rule{0pt}{2.4ex}
	(1-p_a-p_b+2 p_ap_b)(I+\sigma_z) + 
	(p_a+p_b-2 p_ap_b)(I-\sigma_z)\right]/2
\nonumber
\\
	& + & \sin\theta \sigma_x \otimes \left[\rule{0pt}{2.4ex} 
	(1-p_a-p_b) (I+\sigma_z) + (-p_a+p_b) (I-\sigma_z) \right]/2
\\
	&=& \left[\rule{0pt}{2.4ex} \cos\theta (1-p_a-p_b+2 p_a p_b) \sigma_z
	+ \sin\theta (1-p_a-p_b) \sigma_x \right] \otimes (I+\sigma_z)/2
\nonumber
\\
	& + & \left[\rule{0pt}{2.4ex} \cos\theta (p_a+p_b-2 p_ap_b) \sigma_z 
	+ \sin\theta (-p_a+p_b) \sigma_x \right] \otimes (I-\sigma_z)/2
\label{eq:codedout}
\,.
\end{eqnarray}

\end{onecolumn}

\begin{twocolumn}

\noindent {\bf Alternative understanding of the code}

The operation of the code can be further understood using the above mixed
state description.  Without coding, the input qubit is stored in the terms
$\sigma_z \otimes I$ and $\sigma_x \otimes I$ which decay at different rates
and distort the state asymmetrically.  With coding, the input qubit is stored
in the terms $(\sigma_z \otimes \sigma_z + \sigma_y \otimes \sigma_x)$ and
$(\sigma_x \otimes I + I \otimes \sigma_y)$ which decay at the same overall
rate to first order.  The resulting decoded state therefore shrinks
radially up to first order.

When subject to phase damping, the points on the Bloch sphere move inwards
transversely without coding, distorting the shape axisymmetrically
(Fig.~\ref{fig:flows}).  In contrast, the points move somewhat radially with
coding, preserving the spherical shape better.

% FIG 15
\begin{figure}[ht]
\centerline{\mbox{\psfig{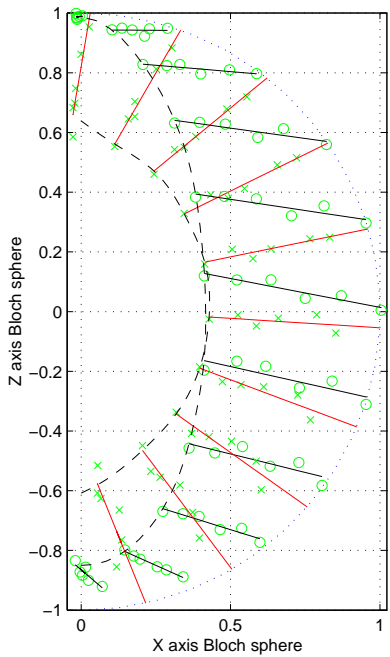}}}
\vspace*{2ex}

\caption{ Experimental data described in Section~\ref{sec:result}.  The solid
	  lines, which are linear fits to the data points, indicate the flow
	  directions of the points on the Bloch sphere subject to phase
	  damping. In the no coding case, the flows are nearly planar, towards
	  the $\hat{z}$-axis, whereas in the encoded case, the flows are
	  nearly radial, towards the center.}
\label{fig:flows}
\end{figure}

%%%%%%%%%%%%%%%%%%%%%%%%%%%%%%%%%%%%%%%%%%%%%%%%%%%%%%%%%%%%%%%%%%%%%%%%%%%%%%%
\section{The case of very different T$_2$'s}
\label{sec:difft2}

While the case of equal $T_2$'s is interesting from a theoretical standpoint,
different spins in a molecule typically have quite different $T_2$'s.  To
study the two bit code in this regime, we performed experiments with carbon-13
labelled chloroform dissolved in acetone \cite{chloroform,Chuang97e}.  
All parameters were similar to the sodium formate sample, except
for the relaxation time constants.

In the chloroform experiment, $T_1$'s were 16~s and 18.5~s and $T_2$'s were
7.5~s and 0.35~s for proton and carbon respectively.  Separate experiments
with the ancilla dephasing much slower or faster than the input were performed
by interchanging the roles of proton and carbon.  $T_2^*$'s and $t_d$'s were
as listed in \cite{chcl3t2}.  The ellipticities are shown in
Fig.~\ref{fig:case23}.

% FIG 16
\begin{figure}[ht]
\centerline{\mbox{\psfig{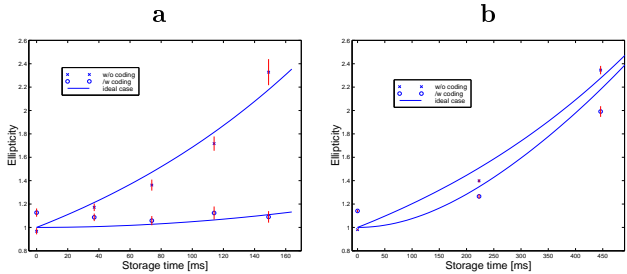}}}
\vspace*{2ex}
\caption{
Ellipticities obtained in the chloroform experiments, with (a) proton and
(b) carbon as the ancilla.  Carbon dephases much faster than proton.  
Error bars represent 95\% confidence level. }
\label{fig:case23}
\end{figure}

From Fig.~\ref{fig:case23}~a, it is apparent that coding almost removes the
distortion entirely when a much better ancilla is available.  The question is,
is coding advantageous over storing in the good ancilla alone?  Theoretically,
coding is always advantageous because the error probability is always reduced
from $p_i$ ($i$ being the input spin) to $p_a p_b$.  Fig.~\ref{fig:case23}~b
shows that experimentally, such improvement is marginal, because the advantage
of coding is offset by the noise introduced.  Therefore, when the $T_2$'s are
very different, the bottle neck is the dephasing of the bad qubit.

%%%%%%%%%%%%%%%%%%%%%%%%%%%%%%%%%%%%%%%%%%%%%%%%%%%%%%%%%%%%%%%%%%%%%%%%%%%%%
\section{Tomography results at major steps}
\label{sec:tomo}

Quantum state tomography~\cite{Chuang97e} is a procedure to reconstruct the
density matrix given a certain set of measurements.  In 2-spin NMR systems,
eight out of the fifteen coefficients, $c_{ij}$, in the Pauli decomposition
are obtainable from the peak integrals, Eq.(\ref{eq:ahigh})-(\ref{eq:blow}).
The remaining 7 parameters can be obtained by repeating the measurement
process with additional readout pulses before acquisition.  These pulses
permute the coefficients $c_{ij}$.  A series of 9 experiments with different
readout pulses is sufficient to reconstruct the complete deviation density
matrix.

We reconstructed the deviation density matrices in the coding experiments.
The results for $\theta = \pi/2$ and $t_d \approx 123$ ms are shown in 
Fig.~\ref{fig:timeseq}.  The ideal matrices were calculated using equations
derived in Appendix~\ref{sec:mixstate}.

\end{twocolumn}

% FIG 17
\begin{onecolumn}
\begin{figure}[ht]
\centerline{\mbox{\psfig{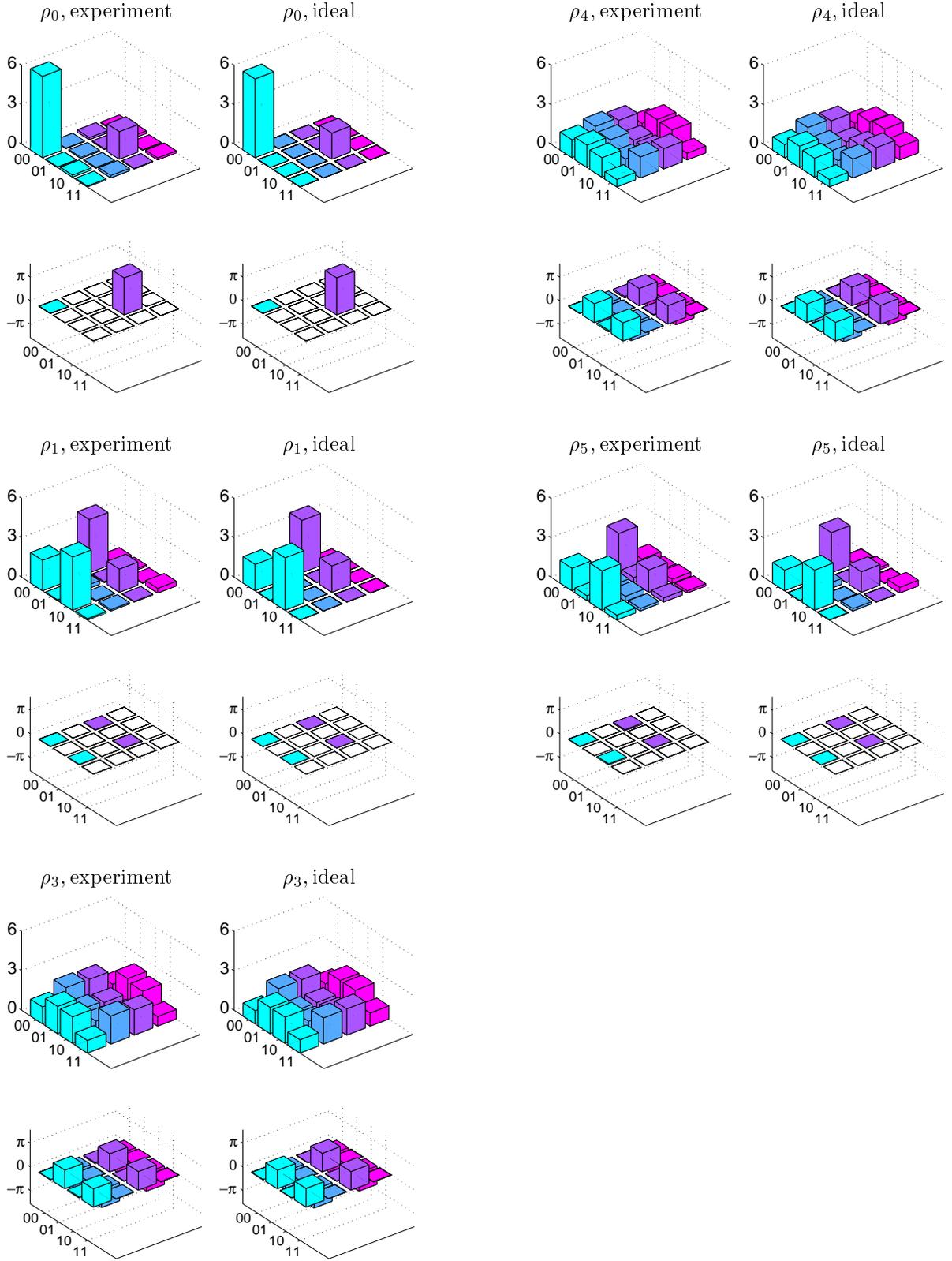}}}
\vspace*{3ex}
\caption{Experimental and ideal density matrices $\rho_0$, $\rho_1$, $\rho_3$,
$\rho_4$ and $\rho_5$.  The basis is as indicated in the diagram.  For each
density matrix, the amplitudes (top) and the phases (bottom) for the
corresponding entries are plotted.  The amplitudes are shown in arbitrary
units and the phases of entries with small amplitudes are omitted.  The
data are taken for a coding experiment with $\theta=\pi/2$ and 
$t_d \approx$ 123 ms (24/$J$).}
\label{fig:timeseq}
\end{figure}
\end{onecolumn}

%%%%%%%%%%%%%%%%%%%%%%%%%%%%%%%%%%%%%%%%%%%%%%%%%%%%%%%%%%%%%%%%%%%%%%%%%%%%%
\end{document}